\documentclass{article}
\usepackage{amsfonts, amssymb, amsmath, amsthm, tocloft,graphicx, caption,subcaption}
\usepackage{slashed}                                                          
\newtheorem{theorem}{Theorem} 	      	      	                              
\newtheorem{lemma}[theorem]{Lemma}     	       	      	      	      	      
\newtheorem{proposition}[theorem]{Proposition} 	      	      	      	      
\newtheorem*{remark}{Remark}                                                  
\newtheorem{definition}[theorem]{Definition} 	      	      	      	      
\newtheorem{assumption}[theorem]{Assumption} 	      	      	      	      

\numberwithin{equation}{section}                                              
\numberwithin{theorem}{section}                                               
\newcommand{\saa}[1]{\marginpar{\tiny {\bf sa:}#1}}

\newcommand{\ol}[1]{\overline{#1}}                                            
\newcommand{\beq}{\begin{equation}}
\newcommand{\eeq}{\end{equation}}
\newcommand{\ul}[1]{\underline{#1}}                                           
\newcommand{\mc}[1]{\mathcal{#1}}                                             
\newcommand{\e}{\epsilon}
\newcommand{\la}{\lambda}
\newcommand{\tla}{\tilde{\lambda}}

\newcommand{\tw}{\tilde{w}}

\newcommand{\tr}{{\rm tr}}

\newcommand{\g}{{\bf g}}
\newcommand{\tgamma}{\tilde{\gamma}}
\newcommand{\lmcN}{\underline{\mathcal{N}}}

\newcommand{\tJ}{{\tilde{J}}}
\newcommand{\up}{\upsilon}
\newcommand{\olg}{{\overline{g}}}
\newcommand{\p}{\prime}

\newcommand{\nasla}{\slashed{\nabla}}                                         

\newcommand{\mind}{/\mspace{-9mu}\gamma}                                      

\newcommand{\tuL}{{\tilde{\underline{L}}}}

\allowdisplaybreaks

\begin{document}

\title{The Penrose inequality on perturbations of the Schwarzschild exterior}
\author{Spyros Alexakis\thanks{University of Toronto, alexakis@math.toronto.edu}}
\date{}
\maketitle

\begin{abstract}
 We prove a version  the Penrose inequality for black hole space-times which are perturbations 
of the Schwarzschild exterior in a slab around a  null  hypersurface 
$\lmcN_0$. $\lmcN_0$ 
  terminates at past null infinity $\mc{I}^-$ and   $\mc{S}_0:=\partial\lmcN_0$ 
is chosen to be a marginally outer trapped sphere. We show that the area of $\mc{S}_0$  yields a lower bound for the Bondi energy of sections of 
 past null infinity, thus also for the total ADM energy.  Our argument is perturbative, 
 and rests on suitably deforming the initial null hypersurface $\lmcN_0$ to one for which 
 the natural ``luminosity'' foliation originally introduced by Hawking  yields a monotonically  increasing Hawking mass, \emph{and} for which the leaves of 
this foliation become asymptotically round. It is to ensure the latter (essential) property 
that we perform the  deformation 
 of the initial null hypersurface $\lmcN_0$.  
\end{abstract}

\tableofcontents

\section{Introduction}
\label{sec-intro}

We prove a version of the Penrose inequality for metrics which are  local perturbations 
of the Schwarzschild exterior, around a (past-directed) shear-free outgoing 
null surface $\lmcN_0$. 

Our method of proof is an extension of the approach sometimes called the ``null\rq{}\rq{} 
Penrose inequality \cite{ludvig-vicker, Mars}, 
and in fact relies on ideas in the PhD thesis of Sauter under the direction of D.~Christodoulou,
 \cite{Sauter}. 
Before stating our result, we 
recall in brief the original formulation of the Penrose inequality and the motivation behind it. 
This will make clear the motivation for proving this inequality for perturbations of the 
Schwarzschild solution. 
\newline

 Penrose proposed  his celebrated inequality 
\cite{P1} as a \emph{test} for what he called the ``establishment view\rq{}\rq{}
on the  evolution of dynamical black holes in the large. For  a single black hole, the view was that the exterior region $(\mc{M}_{\rm ext}, \g)$ 
 should evolve smoothly and eventually settle down, 
due to emission of matter and radiation into the black hole (through the future event 
horizon $\mc{H}^+$) and towards 
future null infinity $\mc{I}^+$. This ``final state'' would  be  
non-radiating, and thus putatively stationary. In vacuum,  such final states are believed to belong 
to  the Kerr family of solutions, \cite{HawkEl, P1}. 

While the mathematical verification of the above appears completely beyond the reach of current 
techniques (and indeed the above scenario rests on some fundamental conjectures of general relativity such as
the weak cosmic censorship), it gives rise to a very rich family of problems. 
For example, the very last assertion is the celebrated 
``black hole uniqueness question\rq{}\rq{}, 
to which much work has been devoted. It has been resolved under the un-desirable assumption of real analyticity by combining the works of many authors--see 
\cite{HawkEl, Carter, Robinson, Chrusc-Cost}
 and references in the last paper; 
more recently the author jointly with A. Ionescu and S. Klainerman 
has proven the result by replacing the assumption of real-analyticity by either that 
 of closeness to the Kerr family of solutions, 
\emph{or} that of small angular momentum on the horizon \cite{aik1, aik2, aik3}. 

 Given the magnitude of the challenge, Penrose proposed an inequality as a \emph{test} of the above prediction: 
Indeed, if the above scenario were true, then the area of any section $\mc{S}$ of the future 
event horizon would have to satisfy:
\beq
\label{Penrose}
\sqrt{\frac{{\rm Area} [\mc{S}]}{16\pi}}\le m_{\rm ADM}.
\eeq
where $m_{\rm ADM}$ stands for the ADM mass of an initial data set for 
$(\mc{M}_{\rm ext}, \g)$. To see that the final state scenario implies this inequality, one 
needs to recall a few
well-known facts: That this inequality holds for the Kerr space-times (in fact equality  is 
achieved for the Schwarzschild solution), that 
 the area of sections of the event horizon $\mc{H}^+$ 
 is increasing towards the future \cite{HawkEl}, while the Bondi mass of sections of $\mc{I}^+$ 
 is decreasing
 towards the future, while initially (at space-like infinity $i^0$) it agrees with the ADM 
mass. 

Since the above formulation pre-supposes the entire event horizon $\mc{H}^+$, 
an alternative version of the above has been  proposed, where \eqref{Penrose} should hold 
for any 
$\mc{S}$  (inside the black hole) which is  
\emph{marginally outer trapped sphere} (MOTS, from now on). Another 
way to then interpret the 
inequality \eqref{Penrose} for such surfaces $\mc{S}$ is that  all asymptoticaly flat vacuum 
space-times containing a MOTS of area $16\pi$, the ADM mass must be 
bounded below by $1$; moreover the bound should be achieved 
precisely by the Schwarzschild solution of mass 1. 
In short, the mass 1 Schwarzschild solution is the global 
minimizer of the 
ADM mass for asymptotically flat space-times containing a MOTS of area $16\pi$. 

Informally, our main result is  a proof that the Schwarzschild solution of mass 1 is a \emph{local}
minimizer of the ADM energy, in the space of regular solutions close to 
Schwarzschild.  
In fact, our result requires only a portion of  a space-time, which is a neighborhood 
of a past-directed outgoing null hypersurface $\lmcN_0$ which emanates from the MOTS 
$\mc{S}_0$ and extends up to past 
null infinity $\mc{I}^-$. It is in that portion of the space-time that we require our metric to be a 
perturbation of the Schwarzschild exterior. 

Before proceeding to state the result precisely, we note the many celebrated results on 
the Penrose inequality in other settings. The Riemannian version of this inequality (corresponding 
to a time-symmetric initial data surface) was proven by Huisken-Ilmanen (for  a connected 
MOTS which in fact is an outermost minimal surface on the initial data surface)
 \cite{HI} and Bray (for a MOTS with possibly many components) \cite{Br}, and has led to many  
 extensions to incorporate charge and angular momentum. 
See \cite{Mars} for  a review of these.  
\newline

While the aforementioned results deal with an asymptoticaly flat (Riemannian) hypersurface in 
a space-time $(\mc{M}, \g)$, 
our point of reference will be null hyper-surfaces in $(\mc{M},\g)$.

 To state our assumptions clearly, recall the  the Schwarzschild metric of mass $m$ in 
  Eddington-Finkelstein coordinates:
\beq
\label{Schwarz}
g_{\rm Schwarz}=-(1-\frac{2m}{r})d\ul{u}^2+2d\ul{u}dr+r^2(d\theta^2
+sin^2\theta d\phi^2).
\eeq
Here  any hyper-surface $\{\ul{u}={\rm Const}, r\ge 2m\}$ is a shear-free
 (rotationally symmetric) past-directed outgoing null 
 hypersurface, which terminates at past null infinity 
 $\mc{I}^-$.\footnote{$\{\ul{u}={\rm Const}, r\ge 2m\}$ is a 3-dimensional smooth
  submanifold. 
We use the term ``hypersurface\rq{}\rq{} or ``null surface\rq{}\rq{} in this paper, with a slight 
abuse of language.}  
 Consider the domain
  $\mc{D}:=\{\ul{u}\in [\ul{u}_0,\ul{u}_0+m), r\ge 2m\}$ 
 in the Schwarzschild space-times. The portion $\{r=2m\}$ of the boundary 
  of this domain 
 lies on the future event horizon $\mc{H}^+$
 of the Schwarzschild exterior. In particular all the spheres 
 $\{r=2m, \ul{u}={\rm Const} \}$ 
 are marginally outer trapped surfaces (the definition of this notion is recalled below).
 
  Our main assumption on our space-time is that the metric $g$ defined over the domain 
  $\mc{D}:= \{\ul{u}\in [\ul{u}_0, \ul{u}_0+m),  r\ge 2m, \theta\in [0,\pi), \phi\in [0,2\pi)\}$
   is a 
 $\mc{C}^4$-perturbation of the Schwarzschild metric over the same domain, where we 
partly  fix the gauge by requiring 
 that the sphere $\mc{S}_0:=\{\ul{u}=\ul{u}_0, r=2m\}$ is still marginally outer 
  trapped  and 
   the hypersurface $\{r=2m\}$ is  still null. 
   (We make the $\mc{C}^4$-closeness  assumption  precise  below).

The inequality that we prove involves the notion of Bondi energy of a section of null infinity, 
compared to the area of a MOTS. We review these notions here: 
\newline

{\bf Sections of Null infinity and the Bondi energy.} We recall past null infinity $\mc{I}^-$ is 
an idealized boundary of the space-time, where past-directed 
null geodesics ``end''. In the asymptotically flat 
 setting that we deal with here, the topology of $\mc{I}^-$ is $\mathbb{R}\times
 \mathbb{S}^2$. In the  coordinates introduced above for the Schwarzschild exterior 
 (and also in the perturbed Schwarzschild space-times we will consider), we can endow
  $\mc{I}^-$ with the coordinates $\ul{u}, \phi,\theta$. 
  
  A $\mc{C}^2$ \emph{section} $\mc{S}_\infty:= \{\ul{u}=f(\phi,\theta)\}$ of $\mc{I}^-$ 
  can be seen as the boundary at infinity of an (incomplete) null surface $\lmcN$. We
   schematically write $\mc{S}_\infty=\partial_\infty\lmcN$. Now, consider any 1-parameter
    family of spheres $\mc{S}_t, t\ge 1$ foliating $\lmcN$ and converging (in a topological 
    sense) 
   towards $\mc{S}_\infty$. A (suitably regular) such 
    foliation $\{\mc{S}_t\}$ is thought of as a reference frame 
   relative to which certain natural quantities at $\mc{S}_\infty$ are defined. In particular, 
recalling the Hawking mass of any such sphere via \eqref{hawk.mass} below,  let: 
   
   \begin{definition}
   \label{Bondi}
   The Bondi energy of $\mc{S}_\infty:=\partial_\infty\lmcN$ relative to the foliation (reference 
   frame)  $\mc{S}_t$ is defined to be :
   
   \[
   E_B:=lim_{t\to\infty} m_{\rm Hawk} [\mc{S}_t],
   \]
   
   provided that: 
   \[
   \lim_{t\to\infty} \big{(}\frac{{\rm Area}[\mc{S}_t]}{4\pi}\mc{K}[\mc{S}_t]\big{)}=1.
   \]
(The limit here makes sense by pushing forward the metrics from $\mc{S}_t$ to $\mc{S}_\infty$, via a natural map that identifies the point on $\mc{S}_t$ with the one 
on $\mc{S}_\infty$ 
that lies on the same null generator of $\lmcN$).
   \end{definition} 
We note that the \emph{asymptotic roundness} required above is \emph{necessary} for the 
limit of the Hawking masses to correspond to the Bondi Energy of 
$\mc{S}_\infty$ (relative to 
the foliation $\mc{S}_t,t\ge 1$). We also remark that the Bondi energy corresponds to the
 time-component of the Bondi-Sachs  energy-momentum 4-vector $(E_B, \vec{P}_B)$ 
 cf Chapter 9.9 in \cite{PR}, relative to the reference frame given by $\mc{S}_t, t\ge 1$. 
 
  The Bondi \emph{mass} is defined to be the Minkowski length of this vector:
  
  \[
  m_B[\mc{S}_\infty]= \sqrt{(E_B)^2-|\vec{P}_B|^2}.
  \]
 This is in fact invariant under the changes of reference frame considered above. We refer to \cite{PR}
 Chapter 9.9 for the definition of these notions. We note that the theorem we prove here 
 yields a lower bound for the Bondi energy, rather than the Bondi mass. In fact, we believe that
 the proof can be adapted to cover the case of the Bondi mass also,\footnote{We explain why 
 this should be so in a remark in the next subsection. } since the two notions agree 
 when the linear momentum vanishes (this is the center-of-mass reference frame). 
 
 The reason we do not pursue this here, is that the definition of linear momentum
used  in \cite{PR} uses spinors, and it is not immediately clear to the author how to translate this 
notion 
into the framework of Ricci coefficients used here. 
\newline

 {\bf Marginally outer trapped spheres:} 
 \begin{definition}
In an asymptoticaly flat space-time $(\mc{M}, \g)$, 
 a 2-sphere  $\mc{S}\subset \mc{M}$ is called marginally outer trapped if, letting 
 $tr\chi^L[\mc{S}]$ 
 be its null expansion relative to a 
 future-directed outgoing normal null vector field $L$,  and also 
 $tr\chi^{\ul{L}}[\mc{S}]$  its null expansion relative to the future-directed
  incoming null normal 
 vector field $\ul{L}$, then for all points $P\in\mc{S}$:
 \[
 tr\chi^L[\mc{S}](P)=0, tr\chi^{\ul{L}}[\mc{S}](P)<0.
 \]
 
   \end{definition}

   Our result is the following: 
   
    \begin{theorem}
    \label{the-result}
   Consider any vacuum Einstein metric over $\mc{D}:=\{ \ul{u}\in [\ul{u}_0, \ul{u}_0+m), 
   r\ge 2m, (\phi,\theta)\in\mathbb{S}^2 \}$ as above which is a perturbation 
   of the Schwarzschild
   metric of mass $m$ over $\mc{D}$ (measured to be $\delta$-close, 
   in a suitable norm that we introduce below),
    and such that the surface $\mc{S}_0:=\{\ul{u}=\ul{u}_0, r=2m\}$ is 
    marginally outer trapped.
   
   Then if $\delta\ll m$ there exists  a  perturbation $\mc{S}'\subset \mc{D}$ of 
   $\mc{S}_0$, which is also
    marginally outer trapped, and 
   such that
\begin{itemize}
\item ${\rm Area}[\mc{S}\rq{}]\ge {\rm Area}[\mc{S}]$. 
\item The past-directed outgoing null surface $\lmcN_{\mc{S}^\p}$ 
emanating from $\mc{S}^\p$ is smooth, terminates 
   at a cut $\mc{S}^\infty\subset\mc{I}^-$, and moreover the following Penrose 
   inequality holds:
   
   \beq
   E_{\rm B}[\mc{S}^\infty]\ge\sqrt{\frac{{\rm Area}[\mc{S}']}{16\pi}},
   \eeq
where $E_{\rm B}[\mc{S}^\infty]$ stands for the Bondi energy on the
 (asymptotically round) sphere $\mc{S}^\infty$ on $\lmcN_{\mc{S}'}$, associated with 
the luminosity  foliation on $\lmcN_{\mc{S}^\p}$. (The latter foliation  is recalled below) 
\end{itemize}
Furthermore, equality holds in the first inequality if and only if 
$\mc{S}^\p\subset 
\{r=2m\}$, and $tr\chi^L=0$ on $\{r=2m\}$ between $\mc{S}, \mc{S}^\p$.
 Equality holds in the second inequality  if and only if 
 $\lmcN_{\mc{S}'}$ is isometric (intrinsically and extrinsically)
 to a spherically symmetric outgoing null surface $\{\ul{u}={\rm Const}\}$ in a 
 Schwarzschild space-time.
   \end{theorem}

\begin{figure} \label{non-char}
\centering
\includegraphics[height=200pt]{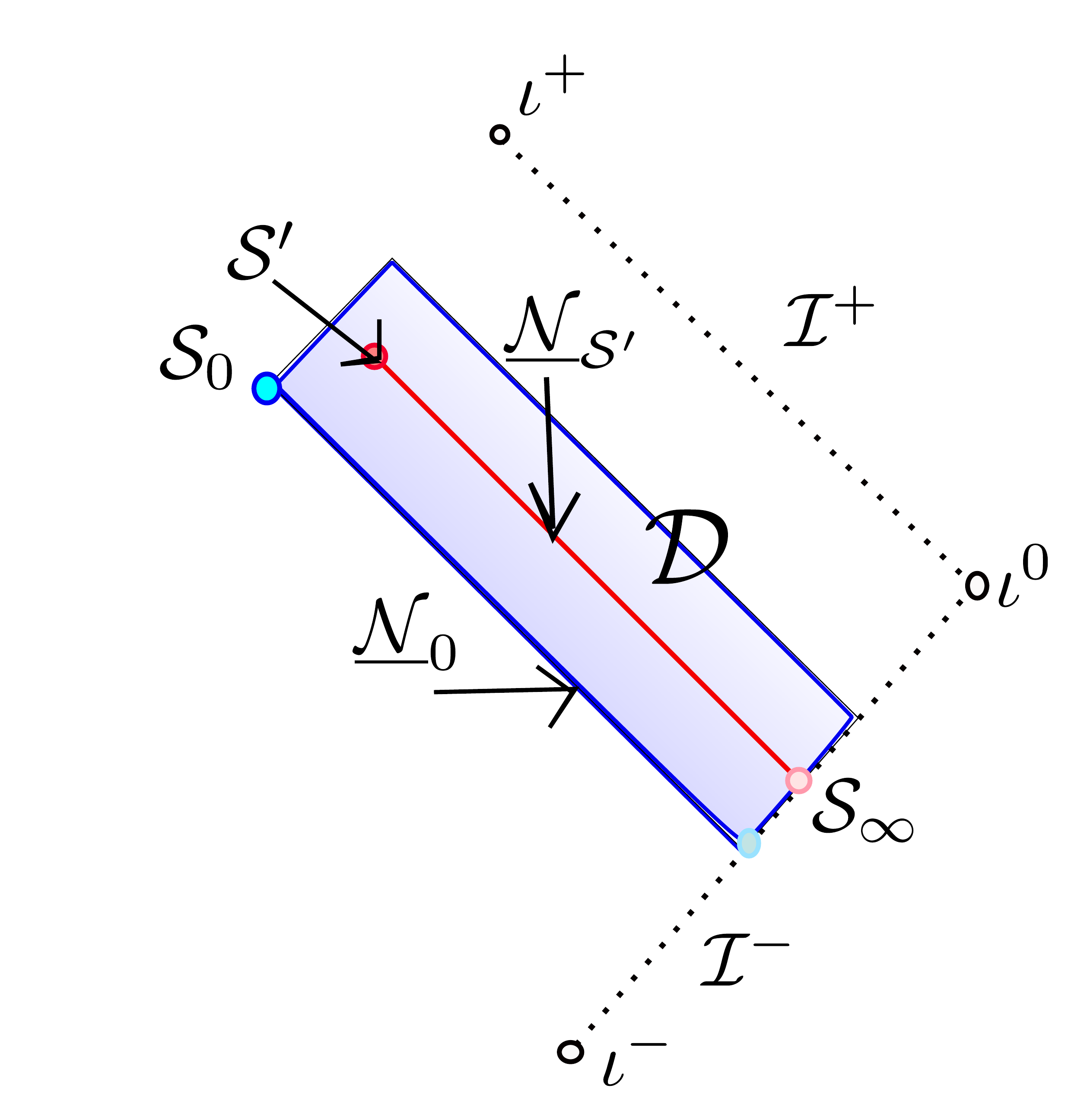}
\label{fig.annuli}
\caption{The Penrose diagram of the domain $\mc{D}$, the ``old''  MOTS $\mc{S}$ and the ``new'' MOTS $\mc{S}^\p$, 
and the smooth outgoing null surfaces $\lmcN_0, \lmcN_{\mc{S}^\p}$.}
\end{figure}

    \subsection{Outline of the paper.}
    
     Our proof rests on a perturbative argument. We mainly seek to exploit the  
evolution equation of the Hawking mass under a particular law of motion (introduced in \cite{H})  
on a \emph{fixed}, 
\emph{smooth}, outgoing null surface together with the possibility of perturbing the underlying 
null surface itself. We note that the possibility of varying the underlying null hypersurface 
as a possible approach method towards deriving the Penrose inequality 
was raised in Chapter 8 in \cite{Sauter}.

The broad strategy is based on two observations: 

\begin{itemize}
\item On the Schwarzschild space-time \emph{and},  on small perturbations of the
 Schwarzschild space-time, the Hawking mass is \emph{increasing} along ``nearly'' shear-free 
null hypersurfaces $\lmcN_0$ that emanate from a MOTS, when $\lmcN_0$ is foliated by a 
``luminosity parameter'', originally introduced in \cite{H}, as we recall below. 
However, as observed in \cite{Sauter}, 
the corresponding leaves of the foliation may \emph{fail} to become asymptotically round.

\item Under the closeness to Schwarschild assumption, we can perturb the underlying 
null hypersurface $\lmcN_0$ towards the future in order to induce a small 
conformal deformation of the  metric ``at infinity'' associated with such a foliation. 
In fact (after renormalizing by the area), we can 
achieve \emph{any} small conformal deformation. Moreover, all 
such new null hypersurfaces $\lmcN'$ emanate from MOTSs $\mc{S}'$ with 
${\rm Area}[\mc{S}']\ge {\rm Area}[\mc{S}]$.
\end{itemize}
The proof is then finished by invoking the implicit function theorem, since by the ``closeness to 
Schwarzschild'' assumption, the (renormalized) sphere at infinity $\mc{S}_\infty$ of the original 
is ``nearly round'', thus it can be made exactly round by a small conformal deformation.  

\begin{remark}
As explained, our proof below in fact shows that \emph{all} small, area-normalized,
 conformal transformations 
of the metric on the sphere at infinity $\mc{S}_\infty$ can be achieved by small 
perturbations of 
$\lmcN_0$. In particular, we believe that we can find a nearby surface $\lmcN'$ for which 
the \emph{linear momentum} of the (perturbed)  sphere at infinity vanishes, thus we obtain a 
lower bound for the Bondi mass, and not just the Bondi energy. However we do not pursue  
  this here. 
\end{remark}

{\it Section Outline:}
In section \ref{sec-assumptions} we state the assumptions of closeness to Schwarzschild 
precisely. We 
also recall the Ricci coefficients associated to a foliated null surface $\lmcN$, and (certain of)  the null
 structure equations which link these to components of the ambient Weyl curvature.  
We also set up the framework for the analysis of perturbations
of our given null surface  $\lmcN_0$. In section \ref{sec-MOTS} we study the ``nearby'' MOTSs
to $\mc{S}_0$, when we deform $\mc{S}_0$ towards the future. 

In section \ref{sec-monotonicity} 
we study the asymptotic behaviour of the expansion $\ul{\chi}$ 
of any smooth fixed null surface $\lmcN$, and moreover the variational behaviour of this relative 
to perturbations of $\lmcN$. We also recall the luminosity foliations and the monotonicity of the 
Hawking mass evolving along these foliations. We then show that the luminosity foliations
 asymptotically agree with  \emph{affine} foliations to
  a sufficiently fast rate, so that  the Gauss curvatures and Bondi energies associated to the 
  two foliations  agree. 
  This then enables us to \emph{replace} the study of luminosity foliations by affine ones. 

 In particular in section \ref{sec-variations} the first variation of the luminosity foliations is 
 captured by ``standard'' Jacobi fields. We then 
 note (relying on some calculations in \cite{AlexShao3}) that the effect of a variation of 
 $\lmcN_0$ 
 on the Gauss curvature at infinity is captured by a conformal transformation of the underlying 
 metric over the sphere at infinity $\mc{S}_\infty$. 
In the latter half of that 
section  we  derive the solution of the relevant Jacobi fields. In conclusion, the first 
variation of the Gauss curvature is captured in the composition of two second-order operators,
$\mc{F}\circ\mc{L}$, where $\mc{L}$ is a perturbation of the Laplace-Beltrami operator 
on the metric $\gamma_0:=\g|_{\mc{S}_0}$, and $\mc{F}$ is the operator 
$\Delta_{\gamma^\infty}+2\mc{K}[\gamma^\infty]$, for $\gamma^\infty$ being the 
metric at infinity associated with the luminosity foliation on $\lmcN_0$ and 
$\mc{K}[\gamma^\infty]$ its Gauss curvature. 
The proof is then completed in section \ref{sec-inverse} by an application of the implicit
function theorem. 
\newline

{\bf Acknowledgements:} The author thanks Rafe Mazzeo for helpful conversations. Some ideas 
here also go back to the author's joint work with A. Shao in \cite{AlexShao3}. This research was
 partially supported by grants 488916 and
489103 from NSERC and an ERA Ontario grant.

\section{Assumptions and Background.}
\label{sec-assumptions}

We state the assumptions on the (local) closeness of our space-time to the Schwarzschild 
solution, as well as the norms in which this closeness is measured. 
The assumptions we make concern the metric in the domain $\mc{D}$, and the curvature 
components 
 and their variational properties on a suitable family of smooth null hypersurfaces 
contained inside $\mc{D}$. (This is the family in which we perform the variations of 
$\lmcN_0$). As we note below, 
one expects that one does not really need to assume regularity on the whole domain $\mc{D}$. 
Rather, what we assume here should be derivable assuming data on the initial $\lmcN_0$, 
and a suitable portion $\ul{u}\in [\ul{u}_0, \ul{u}_0+m)$ 
of $\mc{I}^-$ only. Finally, we note that the number of derivatives
(of the various geometric quantities) can be dropped. Yet any such weakening of the 
assumptions is not in the scope of this paper, and would  lengthen it substantially. 
The main contribution we wish to make here is to introduce 
the idea of deforming the null surfaces under consideration, as a possibly useful method for
 inequalities in general relativity, via an ODE analysis of the null structure equations.

\subsection{Closeness to Schwarzschild and regularity assumptions.}

  In order to state the closeness assumption precisely from $(\mc{M}, \g)$, we need to 
  introduce the parameters that capture the closeness of our space-time 
  to the Schwarzschild background. 
  \newline

{\bf The Scharzschild metric:}   Recall the form \eqref{Schwarz} of the Schwarzschild metric $g_{\rm Schwarz}$. 
  In particular note that $r$ is both an affine parameter on each null surface 
  $\ul{u}={\rm Const}$, and an area parameter, in that:
  \[
  {\rm Area}[\{r=B\}\bigcap \{\ul{u}={\rm Const}]\}=4\pi B^2.
  \]
   Consider the normalized null vector fields

  \beq
  \ul{L}=\partial_r, L=2\partial_{\ul{u}}+(1-\frac{2m}{r})\partial_r.
  \eeq
  
Then $g_{\rm Schwarz}(L,\ul{L})=2$ and  the Ricci coefficients of the Schwarzschild metric relative to this pair of null vectors are 

\beq
tr\chi^L=(1-\frac{2m}{r})\frac{2}{r}, tr\ul{\chi}^{\ul{L}}=\frac{2}{r}, \hat{\chi}^L=\hat{\ul{\chi}}^{\ul{L}}=0, \zeta^L=0.
\eeq
  
In particular note that $tr\chi^L>0$ on $\lmcN$ away from $\mc{H}^+=\{r=2m\}$,   and moreover 
\beq
\partial_r tr\chi|_{\mc{H}^+}=(2m)^{-1}>0.
\eeq
   
{\bf Coordinates for the perturbed metric:} The metrics we will be considering will be 
perturbations of a Schwarzschild metric of mass $m>0$ over a domain $\mc{D}$. 
We will be using a label $u$ for the outgoing past-directed parameter instead of $r$; this is 
since it will no longer correspond with the area parameter $r$. 
We consider a metric $g$ over 
$\mc{D}:=\{\ul{u}\in [\ul{u}_0,\ul{u}_0+m), u\ge 2m, \phi\in [0,2\pi), 
\theta \in [0,\pi) \}$. 
We let $\mc{S}_0:=\{u=2m, \ul{u}=\ul{u}_0\}$.
(Below we will always   be considering two coordinate systems to cover this 
sphere;
 all bounds will be assumed to hold with respect to either of the coordinate
  systems). 
The coordinates are normalized so that $u$ is an affine parameter on each surface 
$\{\ul{u}=\rm Const\}$, and
 the level set $\mc{N}[\mc{S}]:=\{u=2m\}$ is an outgoing null hypersurface,
  and moreover $\ul{u}$ 
 is an affine parameter on $\mc{N}[\mc{S}_0]$.
 The coordinates $\phi,\theta$ are fixed on the initial sphere $\mc{S}_0$ and are then 
 extended to be constant on the generators of $\mc{N}[\mc{S}_0]$, and then \emph{again}
 extended to be constant on each of the null generators of each $\{\ul{u}=\rm Const\}$. 
 
 Note that this condition fixes
  the coordinates uniquely, up to normalizing $\partial_{\ul{u}},\partial_u$ on 
the initial sphere, $\mc{S}_0$. We normalize so that letting 
$\ul{L}:=\partial_u$ on that sphere $tr\ul{\chi}^{\ul{L}}=2$ and 
$g(\ul{L}, L)=g(\ul{L},\partial_{\ul{u}})=2$.
\newline

We will find it convenient
to introduce a canonical frame on \emph{any} smooth past-directed outgoing 
null surface $\lmcN$ emanating from 
a sphere $\mc{S}$; the 
frame is uniquely determined after we choose (two) coordinate systems that cover $\mc{S}$, 
and an affine parameter on $\lmcN$. 

\begin{definition}
\label{frame-constr}
Given coordinates $\phi,\theta$ on $\mc{S}$, 
we extend them to be constant along the null generators of $\lmcN$.
Given an affine parameter $\la$ on $\lmcN$, with $\la=1$ on 
$\mc{S}=\partial\lmcN$, we let 
$\Phi:=\la^{-1}\partial_\phi, \Theta:=\la^{-1}\partial_\theta$. 
We let $e^1:=\Phi, e^2:=\Theta$. 

We also consider a vector field $L$ which is defined to be 
 future-directed and null, and moreover satisfies:
\beq
g(e^1, L)=g(e^2,L)=g(L,\ul{L})=2.
\eeq
\end{definition}

We will be measuring the various natural tensor fields over a null surface  $\lmcN$  
with respect to  the 
above frame. We introduce a measure of smoothness of such tensor fields:

\begin{definition}
\label{O}
We say that a function $f$ defined over any smooth infinite null surface $\lmcN$ with an affine parameter $\la$ 
belongs to the class 
$O^\delta_2(\la^{-K})$ if in the coordinate system $\{\la,\phi^1:=\phi, \phi^2:=\theta\}$ 
on $\lmcN$ we have $|\la^Kf|\le\delta, 
|\partial_{\phi^i}\la^Kf|\le \delta$, 
$|\partial^2_{\phi^i\phi^j} \la^K f|\le \delta$. The classes $O^\delta_1, O^\delta$ are defined
analogously, by not requiring the last (respectively, the two last) estimates above to hold. 

We also considering any 1-parameter family of tensor fields $t_{ab\dots c}$ over the spheres 
$\mc{S}_\la$ that foliate $\lmcN_u$. 
(Thus the indices take
 values among $e^1, e^2$). We say that $t\in O^\delta_2(\la^{-K})$ if and 
 only if any component $t(e_{i_1}\dots e_{i_d})$ in this frame belongs to 
 $O^\delta_2(\la^{-K})$.
\end{definition}
The parameter $\delta>0$ will be our basic measure of closeness 
to the Schwarzschild metric. We {\bf choose} $0<\delta\ll m$. 
\newline

We can now state our assumptions at the level of the metric, the connection coefficients and 
of the curvature tensor. 
Note that for the level sets $\lmcN_{\ul{u}}$ of the optical function $\ul{u}$ introduced above, 
we have chosen a \emph{preferred} affine parameter, which we have denoted by $u$. 
The assumptions will be stated in terms of  that  $u$:

\emph{Assumptions on the metric:} In the coordinate system above over $\mc{D}$, the metric 
$\g$ 
takes the following form, subject to the convention that 
components that are not specifically written out are zero: 

\beq
\begin{split}
&g=(2+O^\delta(u^{-1}))d\ul{u}du -(1-\frac{2m}{u}+O^\delta(u^{-1}))d\ul{u}d\ul{u}+
O^\delta(u^{-1})d\phi d\ul{u}+
O^\delta(u^{-1})d\theta d\ul{u}
\\&+u^2
\sum_{\phi,\theta}g_{\phi\theta}d\phi d\theta,
\end{split}
\eeq
 and the components $g_{\phi\phi}, g_{\phi\theta},g_{\theta\theta}$ are assumed to satisfy  in both coordinate systems:
\beq
|g_{\phi\phi}-(g_{\mathbb{S}^2})_{\phi\phi}|+ |g_{\theta\theta}-
(g_{\mathbb{S}^2})_{\theta\theta}| +|g_{\phi\theta}-
(g_{\mathbb{S}^2})_{\phi\theta}|\le \delta.
\eeq
where $g_{\mathbb{S}^2}$ stands for the standard round metric on the unit sphere 
$\mathbb{S}^2$, with respect to the chosen coordinates $\phi,\theta$.

{\bf Connection coefficients on $\mc{S}_0$:} We assume that

\beq
\label{connectionS0}
\begin{split}
&tr\ul{\chi}^{\ul{L}}[\mc{S}_0]=m^{-1}, \sum_{i=0}^2| \partial^{(i)}\hat{\ul{\chi}}^{\ul{L}}|
\le \delta, 
\sum_{i=0}^2| \partial^{(i)}\hat{\ul{\chi}}^{\ul{L}}|\le \delta, t\chi^L[\mc{S}_0]=0, 
\sum_{i=0}^2| \partial^{(i)}\hat{\chi}^{L}|
\le \delta
\\& \sum_{i=0}^2| \partial^{(i)}\zeta^{L}|\le\delta.
\end{split}
\eeq



{\bf Curvature bounds:} 
Finally, the curvature components are assumed to decay 
towards $\mc{I}^-$ at rates that are consistent with those  
derived by Christodoulou and Klainerman in the stability of the Minkowski 
space-time, \cite{CK}.\footnote{Note that in \cite{CK} the bounds are derived towards 
$\mc{I}^+$.} 
It is necessary, however, to strengthen that assuming that up to 
\emph{two spherical} and one \emph{transverse}\footnote{The transverse directions are
$L$, $\ul{L}$} derivatives of our curvature components also satisfy suitable decay properties. 
(The latter can be seen as \emph{strengthenings} of the decay derived in \cite{CK}, 
which are however entirely consistent with those results).

 To phrase this precisely, recall the independent components of the Weyl curvature in a suitable frame: 
 Consider the level spheres $\mc{S}_{u}\subset\lmcN_{\ul{u}}$ of $u$ on
  $\lmcN_{\ul{u}}$. Consider the frame 
$\Phi,\Theta$ (Definition \ref{frame-constr}) on these level spheres,  and let $L$ be the future-directed
 null normal vector field to the spheres $\mc{S}_u$, normalized so that $g(L,\ul{L})=2$. 

Then, letting the indices $a,b$ below  take values among the vectors $\Phi,\Theta$ we 
require
that for some $\delta>0$ and \emph{all} $\ul{u}\in [\ul{u}_0,\ul{u}_0+m)$, 
$u\ge 2m$:

 \beq
 \label{curv.decay}
 \begin{split}
& \alpha_{ab}:=R_{LaLb}=O^\delta_2(\frac{1}{u}), \beta_a:=R_{L\ul{L}La}=O^\delta_2(\frac{1}{u^2}), 
 \rho:=\frac{1}{4}R_{L\ul{L}L\ul{L}}=-\frac{2m}{u^3}+O^\delta_2(u^{-3})
 \\&\sigma:=\frac{1}{4}R_{L12\ul{L}}=O^\delta_2(u^{-3}), \ul{\beta}_a:=
 R_{\ul{L}L\ul{L}a}=O^\delta_2(u^{-3-\epsilon}), \ul{\alpha}_{ab}:=R_{\ul{L}a\ul{L}b}=
 O^\delta_2(u^{-3-\epsilon}),
\end{split}
 \eeq
where the last two equations are assumed to hold for some $\epsilon>0$.\footnote{Note that
in \cite{CK} these bounds where derived for $\epsilon=\frac{1}{2}$. 
Any $\epsilon>0$ is sufficient for our argument here. }  
 
 We also assume the same results for the $L$-derivatives  and $\ul{L}$-derivatives 
 of the Weyl curvature components,
 again for \emph{all} $\ul{u}\in [\ul{u}_0,\ul{u}_0+m)$, 
$u\ge 2M$:

 \beq
 \label{L.curv.decay}
 \begin{split}
& \nabla_L\alpha_{ab}=O^\delta_2(\frac{1}{u}), \nabla_L\beta_a=
O^\delta_2(\frac{1}{u^2}), 
 \nabla_L\rho=O^\delta_2(u^{-3})
 \\&\nabla_L\sigma=O^\delta_2(u^{-3}), \nabla_L\ul{\beta}_a=
 O^\delta_2(u^{-3-\epsilon}), \nabla_L\ul{\alpha}_{ab}=
 O^\delta_2(u^{-3-\epsilon}).
\end{split}
 \eeq

 \beq
 \label{ulL.curv.decay}
 \begin{split}
& \nabla_{\ul{L}}\alpha_{ab}=O^\delta_2(\frac{1}{u^2}), \nabla_{\ul{L}}\beta_a=
O^\delta_2(\frac{1}{u^3}), 
 \nabla_{\ul{L}}\rho=O^m_2(u^{-4})
 \\&\nabla_{\ul{L}}\sigma=O^\delta_2(u^{-4}), \nabla_{\ul{L}}\ul{\beta}_a=
 O^\delta_2(u^{-4-\epsilon}), \nabla_{\ul{L}}\ul{\alpha}_{ab}=
 O^\delta_2(u^{-4-\epsilon}).
\end{split}
 \eeq

{\bf Variations of null surfaces:} As our theorem is proven by a perturbation argument, 
we now introduce the space of variations
of our past-directed outgoing null surface $\lmcN_0$ which emanates from $\mc{S}_0$.

Considering any smooth positive function $\omega(\phi,\theta)$ over $\mc{S}_0$, 
we let:

\beq
\mc{S}_\omega:=\{\ul{u}=\omega(\phi,\theta)\}\subset \mc{N}[\mc{S}_0].
\eeq

 \begin{definition}
 \label{Nomega}
Let $\ul{L}_{\omega}$ be the unique past-directed and outgoing null vector field 
 normal to $\mc{S}_{\omega}$ and normalized so that 
 $g(L,\ul{L}_{\omega})=2$. 
   Let $\lmcN_{\omega}$
 be the null surface emanating from  $\ul{L}_{\omega}$. We extend $\ul{L}_\omega$
 to an affine vector field: $\nabla_{\ul{L}_\omega}\ul{L}_\omega=0$. 
 Also let $\la_{\omega}$
 be the corresponding affine parameter with $\la_\omega=1$ on $\mc{S}_\omega$.
\end{definition}

The next assumption asserts that the decay assumptions
for the curvature components and Ricci coefficients on all $\lmcN_{\ul{u}}$ 
\emph{persist} under small deformations of $\lmcN_{\ul{u}}$, for all 
$\ul{u}\in [\ul{u}_0,\ul{u}_0+m)$.

For any function $\omega\ge 0$ with $||\omega||_{\mc{C}^2}\le 10^{-1}m$ 
we consider the
 components $\ul{\alpha},\dots, \alpha$ using the vector field $\ul{L}_\omega$, 
and its induced frame on $\lmcN_\omega$. We then 
 assume that the differences of all the relevant curvature 
 components and derivatives thereof between $\lmcN,\lmcN_\omega$ are bounded
 by  $||\omega||_{\mc{C}^2}$. (Recall that by the unique construction of coordinates 
 $\phi,\theta,\la_\omega$ on $\lmcN_\omega$ we have a natural map between 
 $\lmcN,\lmcN_\omega$).
   Specifically:
\begin{assumption}
 For any  $e^a, e^b, a,b\in\{1,2\}$ and any $k,l\in\{1,2\}$ below we assume: 
 \beq
 \label{differences}
 |\alpha^\omega_{ab}-\alpha_{ab}|\le \frac{||\omega||_{\mc{C}^2}}{\lambda}, 
  |\partial_{\phi^k}\alpha^\omega_{ab}-\partial_{\phi^k}\alpha_{ab}|\le 
  \frac{||\omega||_{\mc{C}^2}}{\lambda},  |\partial^2_{\phi^k\phi^l}\alpha^\omega_{ab}-
  \partial^2_{\phi^k\phi^l}\alpha_{ab}|\le \frac{||\omega||_{\mc{C}^2}}{\lambda};
 \eeq
 we moreover assume the analogous bounds for the differences for the derivatives 
 $\nabla_{\ul{L}_\omega}\alpha$ and $\nabla_{L_\omega}\alpha$, and any Ricci coefficient, 
 curvature component, or rotational derivative thereof in 
  \eqref{curv.decay}, 
 \eqref{L.curv.decay}, \eqref{ulL.curv.decay} evaluated against the frames $e^i$ and
  identified via
 the coordinates constructed above. 
\end{assumption}

\begin{remark} The above assumption can in fact be \emph{derived} using  
\eqref{curv.decay}, 
 \eqref{L.curv.decay}, \eqref{ulL.curv.decay} and  by studying the geodesics emanating from 
 $\ul{L}_\omega$
in the coordinate system constructed on $\mc{D}$. However to do this would be somewhat technical and is beyond the scope of this paper. So we prefer to state it as an assumption. 
\end{remark}

Since our argument will be perturbational, we find it convenient to alternatively 
express $\omega=\tau\cdot e^v$, where $\tau$ will be a parameter of variation. 
Specifically:

 \begin{definition}
 \label{Nvt}
Consider any 
 function $v\in \mc{C}^2[\mc{S}_0]$, $||v||_{\mc{C}^2(\mc{S}_0)}\le 10^{-1}m$ and any number 
$\tau\in(0,1)$, and  let:

\beq
\mc{S}_{v,\tau}:=\{\ul{u}=e^v\tau \}\subset\mc{N}[\mc{S}_0].
\eeq
 We also let $\ul{L}_{v,\tau}$ be the unique past-directed and outgoing null vector field 
 normal to $\mc{S}_{v,\tau}$ and normalized so that 
 $g(L,\ul{L}_{v,\tau})=2$. 

  Let $\lmcN_{v,\tau}$
 be the null surface emanating from  $\ul{L}_{v,\tau}$. Also let $\la_{v,\tau}$
 be the affine parameter on $\lmcN_{v,\tau}$ generated by  
 $\ul{L}_{v,\tau}$, normalized so that $\la_{v,\tau}=1$ on $\mc{S}_{v,\tau}$.
\end{definition}

Next, we will be studying the perturbational properties of the 
 parametrized null surfaces based on the geodesics that emanate from 
 $\mc{S}_{v,\tau}$ in the direction of the null vector field 
 $\ul{L}_{v,\tau}$.

A key in proving our theorem will be in equipping each of the null surfaces $\lmcN_{v,\tau}$ with a suitable foliation. As we will see, it 
is sufficient for our purposes to restrict attention to \emph{affine} foliations of our surfaces $\lmcN_{v,\tau}$. The additional
regularity  assumption 
is then that the variation of the Gauss curvatures of the leaves of such affine foliations is captured to a sufficient degree by the 
Jacobi fields which encode the variations. To make this precise, we introduce the space of variations 
 of affinely parameterized null surfaces.

For  a given $v\in W^{4,p}(\mc{S}_0)$ (for any fixed $p>2$ from now on),
 consider any 1-parameter family of smooth null 
surfaces
$\lmcN_{v,\tau}$. Each of
 these surfaces can be identified with a family of affinelly parametrized 
null geodesics that rule them. In other words, consider any sphere 
$\mc{S}_{v,\tau}\subset\mc{N}[\mc{S}_0]$, 
and a smooth family of affinely parametrized null geodesics 
$\gamma_{v,\tau}(s,Q)$ for $Q\in \mc{S}_{v,\tau}$ ($s$ is the affine parameter);
then the null surface $\lmcN_{v,\tau}$, as long as it is smooth
 can be seen simply as the union of 
the null geodesics $\gamma_{v,\tau}(s,Q)$:
\beq
\lmcN_{v,\tau}=\bigcup_{s\in\mathbb{R}_+,Q\in\mc{S}_{v,\tau}}\gamma_{v,\tau}(s,Q).
\eeq

Consider a function $f(\tau,\mc{S}_0)$ (i.e. $f(\tau,\phi,\theta)$ in coordinates). 
Note that the function $f\la_{v,\tau}$ is \emph{still} an affine function on $\lmcN_{v,\tau}$.

Clearly, the variation of the null surfaces $\lmcN_{v,\tau}$ \emph{and} the associated affine parameters $f\cdot \la_{v,\tau}$ 
is encoded in Jacobi fields along the 2-parameter family of geodesics  $\gamma(\lambda,Q)\subset\lmcN$.

In particular, we consider the Jacobi fields over $\gamma_0(Q), Q\in\mc{S}_0$:

\beq
\label{Jac}
J^{f,v}_Q(\la):=\frac{d}{d\tau}|_{\tau=0}\gamma_{v,\tau}(f\cdot \lambda_{v,\tau},Q).
\eeq

\begin{definition}
\label{gauss-curvrs}
We let $\mc{S}_{v,\tau}^{B,f}:=\{\la^f_{v,\tau}=B \}\subset\lmcN_{v,\tau}$. 
We let $\mc{K}[\mc{S}_{v,\tau}^{B,f}]$ be the Gauss curvature of that
sphere.\footnote{We think of the Gauss curvature as a function in the coordinates $\phi,\theta$.}
We let $\tilde{\mc{K}}[\mc{S}_{v,\tau}^{B,f}]:= B^2 \mc{K}[\mc{S}_{v,\tau}^{B,f}]$. 
We call $\tilde{\mc{K}}[\mc{S}_{v,\tau}^{B,f}]$ the renormalized Gauss curvature of 
$\mc{S}_{v,\tau}^{B,f}$.
\end{definition}

The final regularity  assumption on our space-time  essentially states that 
the Jacobi fields \eqref{Jac} capture to a sufficient degree the variation of the Gauss curvatures 
 of the \emph{spheres at infinity}  associated with the affinely parametrized 
 $\lmcN_{v,\tau}$ above.
 To make this precise, define 
 
 \begin{definition}
\label{o}
We say that a function $F_{v,\tau}(\phi,\theta,B)$ with $v\in W^{4,p}(\mc{S}_0)$,
 $\tau\in [0,1)$ and $(\phi,\theta)\in\mathbb{S}^2, B\ge 1$   lies in 
$ o(\tau)$ if for the set 
 $\mc{B}(0,10^{-1}m)\subset W^{4,p}(\mc{S}_0)$,\footnote{I.e. the ball of radius $10^{-1}m$  in the 
Banach space $W^{4,p}(\mc{S}_0)$.} we have 
$\tau^{-1}F_{v,\tau}(\phi,\theta,B)\to 0$ as $\tau\to 0$, uniformly  for all 
$v\in \mc{B}(0,10^{-1}m)$,
 $(\phi,\theta)\in\mathbb{S}^2, B\ge 1$. 
We also say that $F_{v,\tau}(\phi,\theta,B)$ lies in $o_2(\tau)$ if $F$, 
$\partial_{\phi^i}F$ and 
$\partial^2_{\phi^i\phi^j}F$  lie in $o(\tau)$. 

 We  say 
that 
$F_{v,\tau}(\phi,\theta,B)\in O(B^{-1})$ if for all $v\in\mc{B}(0,10^{-1}m)\in W^{4,p}(\mc{S}_0)$ 
as above,
 $\tau\in [0,1)$ we have 
$F(v,\tau,B)\le CB^{-1}$ for some  fixed $C>0$. We also say that $F\in O_2(B^{-1})$
if $F$,  $\partial_{\phi^i}F$ and 
$\partial^2_{\phi^i\phi^j}F$  lie in $O(B^{-1})$.
\end{definition}
(Note that Definition \ref{O} deals with functions that depend only on $\phi,\theta, \la$, while 
Definition \ref{o} deals with functionals that also depend on $v\in W^{4,p}, \tau\in [0,1)$). 

Our final regularity assumption is then as follows:
\begin{assumption}
\label{assumption}
Let us consider the frame  $\ul{L},\Phi,\Theta,L$ over $\lmcN_0$
as in Definition \ref{frame-constr}

Consider a Jacobi field $J$ over $\lmcN_0$ 
as in \eqref{Jac} for $v\in W^{4,p},p>2$.\footnote{We suppress $f,v,Q$ for simplicity.}
 Express 
$J$ with respect to the frame $\ul{L}, \Phi,\Theta, L$,
 with components $J^{L}, J^{\ul{L}}$ and $J^\Phi, J^\Theta$. 
Assume that $J^{\Phi}, J^\Theta=O_2(\la^{-1})$. 

Then we assume that the (renormalized) Gauss curvature of the sphere 
$\{f\cdot \la_{v,\tau}=B\}\subset\lmcN_{v,\tau}$ asymptotically agrees with that of the sphere obtained by flowing  by $\tau$ in the direction of the Jacobi field starting from 
  $\lmcN_0$: 

Let $\tilde{\mc{S}}_{v,\tau}$  be the sphere 
$\{f\la_{v,\tau}=B\}\subset\lmcN_{v,\tau}$ and $\mc{S}_{1-{\rm flow}(\tau)}(B)$  
be the sphere that arises from $\{\la_{0}=B\}\subset\lmcN$ by flowing along 
$\ul{L}$ by $J^{\ul{L}}\tau$. Finally let $L^\prime_\tau$ be the null vector field that is normal 
to $\mc{S}_{1-{\rm flow}(\tau)}(B)$, and normalized so that $g(L^\prime_\tau, \ul{L})=1$;  
let $\mc{S}_{2-{\rm flow}(\tau)}(B)$ be the sphere that arises from $\mc{S}_{1-{\rm flow}(\tau)}(B)$
flowing along the $L^\prime_\tau$ by $J^L\cdot \tau$. 
Then:

\beq
B^2{\mc{K}}[\tilde{\mc{S}}_{v,\tau}]=
B^2\mc{K}[\mc{S}_{2-{\rm flow}(\tau)}(B)]+o(\tau) +O(B^{-1}).
\eeq 
\end{assumption}

We remark that the above property is entirely standard in a smooth metric for 
\emph{finite} 
geodesic segments. (See the discussion on Jacobi fields in  \cite{Jost}, for example).   
Thus the assumption here should be seen as 
a regularity assumption 
on the space-time metric near null infinity, in the rotational directions $\Phi,\Theta$ 
and in the transverse direction $L$. One expects that this property of Jacobi fields can 
be derived 
from the  curvature  fall-off assumptions we are making, since it also follows immediately 
when the space-time admits a sufficiently regular conformal compactification (by 
using the aforementioned result on finite geodesic segments as well as the conformal invariance 
of null geodesics).
 However proving 
 this is beyond the scope of this paper, so we state this as an assumption.


\subsection{The geometry of a null surface and the structure equations.}

  The analysis we will perform will require the  use of the null structure equations
  linking the Ricci coefficients of a null surface with the ambient curvature components. 
  We review these equations here. We will be using these equations both 
  for future and past-directed outgoing surfaces $\mc{N}$ and $\lmcN$. 

Let us first consider future-directed null outgoing surfaces $\mc{N}$, and let $L$ be an 
affine vector field along $\mc{N}$. 
Let $\la$ being a corresponding affine parameter and $\ul{L}$
 be the null vector field with $\ul{L}$ normal to the level sets of $\la$ and  normalized so that 
 $g(L,\ul{L})=2$.\footnote{Note that this is the reverse condition compared to the usual
 one, where both $\ul{L}, L$ are future-directed. 
 This is also manifested in some of the null structure equations below.}
 
 Given the Levi-Civita connection $D$ of the space-time metric $g$,
 the Ricci coefficients on $\mc{N}$ for this parameter  are the following:
\begin{itemize}
\item Define the \emph{null second fundamental forms} $\chi, \ul{\chi} $ by
\[ \chi (X, Y) = g ( D_X L, Y ) \text{,} \qquad \ul{\chi} (X, Y) = g ( D_X \ul{L}, Y ) \text{,} \qquad X, Y  \text{.} \]
Since $L$ and $\ul{L}$ are orthogonal to $\mc{S}$, both $\chi$ and $\ul{\chi}$ are symmetric.
The trace and traceless parts of $\chi$ (with respect to $\mind$),
\[ tr \chi = \mind^{ab} \chi_{ab} \text{,} \qquad \hat{\chi} = \chi - \frac{1}{2} (tr \chi) \mind \text{,} \]
are often called the \emph{expansion} and \emph{shear} of $\mc{N}$, respectively.
The same trace-traceless decomposition can also be done for $\ul{\chi}$.

\item Define the \emph{torsion} $\zeta $ by
\[ \zeta (X) = \frac{1}{2} g ( D_X L, \ul{L} ) \text{.} \]
\end{itemize}
 The Ricci coefficients on a given sphere $\mc{S}\subset\mc{N}$ depend on the choice of 
the null pair $\ul{L},L$. When we wish to highlight this dependence below, we will write $\chi^L,\ul{\chi}^{\ul{L}},\zeta^{\ul{L}}$. 


The Ricci and curvature coefficients are related to each other via a family of geometric differential equations, known as the \emph{null structure equations} which we now review.
For details and derivations, see, for example, \cite{CK, kl_rod:cg}.

{\bf Structure equations:} 
We use the connection $\nasla_{\ul{L}}$ which acts on smooth 1-parameter families of vector fields over the level sets of an affine parameter $\la$ as follows: 
\beq
\label{nasla.def}
\nasla_{\ul{L}}X(\phi,\theta,\la_0):={\rm proj}_{\{\la=\la_0\}}\nabla_{\ul{L}}X.
\eeq
In other words, $\nasla_{\ul{L}} X$ is merely the projection of $\nabla_{\ul{L}}X$
onto the level sphere of the affine parameter $\la$. 
The definition extends to tensor fields in the obvious way. We analogously define a
connection $\nasla_L$ on foliated future-directed outgoing null surfaces $\mc{N}$.

Then, the following structure equations hold on $\mc{N}$.

\begin{align}
\label{eq.structure_ev} \nasla_L \chi_{ab} &= - \mind^{cd} \chi_{ac} \chi_{bd} - \alpha_{ab} \text{,} \\
\notag \nasla_L \zeta_a &= 2 \mind^{bc} {\chi}_{ab} \zeta_c - \beta_a \text{,} \\
\notag \nasla_L \ul{\chi}_{ab} &= - ( \nasla_a \zeta_b + \nasla_b \zeta_a ) + \frac{1}{2} \mind^{cd} ( \chi_{ac} \ul{\chi}_{bd} + \chi_{bc} \ul{\chi}_{ad} ) - 2 \zeta_a \zeta_b -
 \rho \mind_{ab} \text{.}
\end{align}
In particular the last equation implies:
\beq
\label{var.trulchi}
\nasla_L \tr\ul{\chi}=\frac{1}{2}\tr\chi\tr\ul{\chi}-2div\zeta-2|\zeta|^2-2\rho+
\hat{\chi}\hat{\ul{\chi}}.
\eeq

An analogous system of Ricci coefficients and structure equations hold for the
 past-directed and outgoing null surfaces $\lmcN$.
Now $\ul{L}$ will be an affine vector field on 
 $\lmcN$ and $\la$ will be the corresponding 
 affine parameter. In this case, we let $L$ be the null 
 vector field that is normal to the level sets of $\la$ and normalized so that $g(\ul{L},L)=2$.

 The Ricci coefficients $\ul{\chi}, \chi$ are then defined as above; we define $\zeta$ in this context to be
 $\zeta(X)=\frac{1}{2}g(D_X\ul{L},L)$. We then have the following evolution equations on $\lmcN$: \begin{align}
\label{eq.structure_ev'} \nasla_{\ul{L}} \ul{\chi}_{ab} &= - \mind^{cd} \ul{\chi}_{ac} \ul{\chi}_{bd} - \ul{\alpha}_{ab} \text{,} \\
\notag \nasla_{\ul{L}} \zeta_a &=  2 \mind^{bc} \ul{\chi}_{ab}
 \zeta_c - \ul{\beta}_a \text{,} \\
\notag \nasla_{\ul{L}} \chi_{ab} &= - ( \nasla_a \zeta_b + \nasla_b \zeta_a ) - \frac{1}{2} \mind^{cd} ( \chi_{ac} \ul{\chi}_{bd} + \chi_{bc} \ul{\chi}_{ad} ) - 2 \zeta_a \zeta_b - \rho \mind_{ab} \text{.}
\end{align}
The last equation implies: 
\beq
\label{var.trchi}
\nasla_{\ul{L}} \tr\chi=-\frac{1}{2}\tr\chi\tr\ul{\chi}-2div\zeta-2|\zeta|^2-2\rho-
\hat{\chi}\hat{\ul{\chi}}
\eeq

\emph{Hawking mass: }
For any space-like 2-sphere $\mc{S}\subset \mc{M}$, we consider any pair of normal vector fields to $\mc{S}$,
 $\ul{L},L$, with $\ul{L}$ and $-L$ being past-directed. Let $\chi(\mc{S}), \ul{\chi}(\mc{S})$ be the two 
 second fundamental forms of $\mc{S}$ relative to these vector fields. Let $tr\chi,tr\ul{\chi}$ be the traces of these. 
 We also let 
 \beq
 \label{r}
 r[\mc{S}]:=\sqrt{\frac{{\rm Area}[\mc{S}]}{4\pi}}
 \eeq
Then the Hawking mass of $\mc{S}$ is defined via:

\beq
\label{hawk.mass}
m_{\rm Hawk}(\mc{S}):=\frac{r}{2}[\mc{S}](1-\frac{1}{16\pi}\int_{\mc{S}}tr\chi tr\ul{\chi}).
\eeq
Recall also the mass aspect function $\mu$:
\beq
\label{maspect}
\mu=\mc{K}-\frac{1}{4}tr\chi tr\ul{\chi}-div\zeta.
\eeq
In view of the Gauss-Bonnet theorem, we readily derive that: 
\beq
\label{Hawk-MA}
\int_{\mc{S}}\mu dV_{\mc{S}}=\frac{8\pi}{r}m_{\rm Hawk}(\mc{S}).
\eeq


\subsection{Transformation laws of Ricci coefficients, and perturbations of 
 null surfaces.}
 \label{transf-AS3}

Recall that $\mc{N}[\mc{S}_0]$ is equipped with an affine parameter $\ul{u}$ 
normalized so that $\ul{u}=\ul{u}_0$ on $\mc{S}_0$ and $L(\ul{u})=1$.
Our proof will require calculating $\ul{\chi}^{\ul{L}_{v,\tau}}[\mc{S}_{v,\tau}]$
up to an error $o_2(\tau)$.

For this we will use certain transformation formulas for the Ricci coefficients 
of affinely parameterized null surfaces under changes of the affine foliation; we refer the reader to \cite{AlexShao3} for the details and derivations of these. 

In order to 
reduce matters to that setting, we consider a new 
 affine 
parameter $\ul{u}^\p$ on $\mc{N}[\mc{S}_0]$ defined via:

\beq
\label{new-affine}
\ul{u}^\prime-1:=e^{-v} (\ul{u}-1).  
\eeq
(Thus $\{\ul{u}=1\}=\{\ul{u}^\p=1\}=\mc{S}_0$, and $\mc{S}_{v,\tau}=\{\ul{u}^\p=
\tau\}$).\footnote{Note that by construction $\ul{L}_{v,\tau}$ is normal to the level sets of 
the  affine parameter on $\mc{N}[\mc{S}_0]$.}
We now invoke formula (2.11) in \cite{AlexShao3} for $\mc{S}_{v,\tau}$ which calculates 
the Ricci coefficients $\ul{\chi}',\zeta^\p,\chi^\p$ \emph{defined relative to the vector field 
$e^v \ul{L}_{v,\tau}$}, in terms of the Ricci coefficients $\ul{\chi}, \zeta,\chi$
defined relative to the original vector fields $\ul{L}$. 
While the primed $\prime$ 
and un-primed tensor fields  
live over different tangent spaces (the level sets of two different affine
parameters), there is a natural identification between these level sets subject to which 
the formulas below make sense; 
essentially we compare the evaluations of these tensor fields against 
their respective coordinate vector fields. We refer the reader to \cite{AlexShao3} regarding this 
(technical) point.

Thus in our setting we calculate, on the spheres $\mc{S}_{v,\tau}$:\footnote{The 
differences in some signs and the absence of $e^{-v}$ compared to
(2.11) in \cite{AlexShao3} are due to the different orientations  of $\ul{L}$ and $\ul{L}_{v,\tau}$, and their different
 scalings by $e^v$ relative to $L, L'$ in (2.11) in \cite{AlexShao3}.}

 \begin{align}
\label{eq.cf_chibar} \ul{\chi}^\prime_{ab} &= \ul{\chi}_{ab} - 2 (\ul{u} - 1)  
\nasla_{ab} v - 2 (\ul{u} - 1)  ( \nasla_a v \cdot \zeta_b + \nasla_b v \cdot \zeta_a ) \\
\notag &\qquad - (\ul{u} - 1)^2 \mind^{cd}  \nasla_c v ( \nasla_d v \cdot \chi_{ab} - 
2 \nasla_a v \cdot \chi_{bd} - 2 \nasla_b v \cdot \chi_{ad} ) \\
\notag &\qquad - 2 (\ul{u} - 1) \nasla_a v \nasla_b v \text{.}
\end{align}
Replacing $\ul{u}-1$ as in \eqref{new-affine}, (recall $\ul{u}^\p-1=\tau$) we find:

\begin{align}
\label{eq.cf_chibar'} \ul{\chi}^\prime_{ab} &= \ul{\chi}_{ab} - 2 \tau  \nasla_{ab} e^v - 2 \tau  ( \nasla_a e^v \cdot \zeta_b + \nasla_b e^v \cdot \zeta_a ) \\
\notag &\qquad - (s - 1)^2 e^{2v}\mind^{cd}  \nasla_c v ( \nasla_d v \cdot \chi_{ab} - 2 \nasla_a v \cdot \chi_{bd} - 2 \nasla_b v \cdot \chi_{ad} ).
\end{align}

 Then, using (\ref{var.trulchi})  we find:

\begin{align}
\label{eq.cf_chibar.trace}  &  tr\ul{\chi}^\prime(\mc{S}_{v,\tau}) =tr\ul{\chi}
(\mc{S}_{v,\tau})
 -2 \tau  \Delta_{\mc{S}_0} e^v - 4 \tau  (\nasla^ae^v  \zeta_a ) \\
\notag &\qquad + \tau^2 e^v\gamma^{cd}  \nasla_c v  \nasla_d v \cdot tr\chi - 4\tau^2 
 \nasla_av\nasla_d  v \cdot \chi^{ad}   
\\&=tr\ul{\chi}(\mc{S}_0)+e^{v}\tau\nabla_Ltr\ul{\chi}(\mc{S}_0)-2 \tau  
\Delta_{\mc{S}_0} e^v - 4 \tau  (  \zeta^a \nabla_ae^v)
+o(\tau),
\end{align}
where $\Delta_{\mc{S}_0}$ is the Laplace-Beltrami operator for the restriction
 $\mind$ of the space-time metric $g$ onto $\mc{S}_0$. 
Thus in particular, we derive that on $\mc{S}_{v,\tau}$:\footnote{The terms 
in the last  line are all evaluated on the initial sphere $\mc{S}_0$. The difference between the
 Ricci coefficients on $\mc{S}_0$ and $\mc{S}_{v,\tau}$ is of order $o(\tau)$, by the 
 evolution 
 equations in the previous subsection.}

\beq
\begin{split}
\label{var1}
&tr\ul{\chi}^\prime[\mc{S}_{v,\tau}]
=tr\ul{\chi}[\mc{S}_0]+\tau
\{[\frac{1}{2}tr\chi[\mc{S}_0] tr\ul{\chi}[\mc{S}_0]-2\rho[\mc{S}_0]+\hat{\chi}[\mc{S}_0]
\cdot\hat{\ul{\chi}}[\mc{S}_0]\\&-2div\zeta[\mc{S}_0] -2|\zeta[\mc{S}_0]|^2]e^v-2   
\Delta_{\mc{S}_0} e^v - 4  (  \zeta[\mc{S}_0]^a \nabla_ae^v)\}+o(\tau).
\end{split}
\eeq

For future reference, we also recall some facts from \cite{AlexShao3} on the 
transformation law of the second fundamental forms $\ul{\chi}$ and  $\chi$ on the level sets of suitable affine parameters $\la$ on $\lmcN$. In
particular, given a $\mc{C}^2$ function $\omega$ over $\mc{S}=\partial\lmcN$, 
we consider the new affine parameter $\la^\p$ defined via:
\beq
\label{two.las}
\la^\p-1=e^\omega(\la-1).
\eeq
(Recall that $\la=1$ on $\mc{S}$), 
we let $\ul{L}^\p$ the associated null vector field, and $L^\p$
the null vector field that is normal to the level sets of $\la^\p$, normalized so that 
$g(\ul{L}^\p, L^\p)=2$. We let 
 $\ul{\chi}^\p, \chi^\p$  be the null second fundamental forms  corresponding to $\ul{L}^\p, L^\p$. Then  (subject to the 
identification of coordinates described in section 2 in \cite{AlexShao3}),  
 $\ul{\chi}^\p,\chi^\p$ evaluated at any point on $\lmcN$ equal: 

\beq
\label{ulchi-trans}
\ul{\chi}^\p_{ab}=e^\omega \ul{\chi}_{ab}
\eeq

 \begin{align}
\label{zeta-trans} \zeta^\prime_a &= \zeta_a + (\la - 1) \mind^{bc} \nasla_b v \cdot \chi_{ac} - \nasla_a v \text{,} \\
\label{chi-trans} {\chi}^\prime_{ab} &= e^{-\omega}\{{\chi}_{ab} - 2 (\la - 1)  
\nasla_{ab} \omega - 2 (\la - 1)  
( \nasla_a \omega \cdot \zeta_b + \nasla_b \omega \cdot \zeta_a ) \\
\notag &\qquad - (\la - 1)^2 \mind^{cd}  \nasla_c \omega ( \nasla_d \omega
 \cdot \ul{\chi}_{ab} - 2 \nasla_a \omega \cdot \ul{\chi}_{bd} - 2 \nasla_b \omega
 \cdot \ul{\chi}_{ad} ) \\
\notag &\qquad - 2 (\la - 1) \nasla_a \omega \nasla_b \omega \}\text{.}
\end{align}
 
An application of these formulas will be towards constructing new MOTSs, off of the 
original MOTS $\mc{S}_0$.

\section{New MOTS off of $\mc{S}_0$.}
\label{sec-MOTS}

 Our aim here is to capture the space of  marginally outer trapped  
 2-spheres nearby the original sphere $\mc{S}_0$, but to its future, (with respect to the 
 direction $L$).
   Recall $\mc{S}_{v,\tau}, \ul{L}_{v,\tau}$ 
from Definition \ref{Nomega} (where $e^v\cdot\tau=\omega$).

 We let $\ul{\chi}_{v,\tau}$ be the
 null second fundamental form on $\lmcN_{v,\tau}$ corresponding to $\ul{L}_{v,\tau}$.
 We then claim:
 
 \begin{lemma}
\label{newMOTS}

 Given any $v\in\mc{C}^2[\mc{S}_0]$ with $||v||_{\mc{C}^2}\le 
 (10)^{-1}m$, then for all $\tau\in [0,1)$
there exists a function $F(v,\tau)\ge 0$ for which  
$\mc{S}'_{v,\tau}:=\{\la_{v,\tau}=F(v,\tau)\}\subset\lmcN_{v,\tau}$ 
is marginally outer trapped (see Figure 2). Furthermore, we claim that:

\beq
\label{new.trulchi}
\begin{split}
&tr\ul{\chi}[\mc{S}'_{v,\tau}]=2+
\tau\{[-2\rho[\mc{S}_0]+\hat{\chi}_{ab}[\mc{S}_0]
\cdot\hat{\ul{\chi}}^{ab}[\mc{S}_0]\\&-2div\zeta[\mc{S}_0] -2|\zeta[\mc{S}_0]|^2]e^v-2   
\Delta_{\mc{S}_0} e^v - 4  (  \zeta^a[\mc{S}_0] \nabla_ae^v)
\\&+ e^v |\hat{\chi}|^2[\mc{S}_0][
-2\rho[\mc{S}_0]-2div\zeta [\mc{S}_0]-2|\zeta[\mc{S}_0]|^2+\hat{\chi}
[\mc{S}_0]\hat{\ul{\chi}}[\mc{S}_0]]^{-1}\big{(}-2
-|\hat{\ul{\chi}}|^2[\mc{S}_0]\big{)} \}
+o_2(\tau)
\end{split}
\eeq
 \end{lemma}
\begin{figure}
   \label{non-char2}
\centering
\includegraphics[width=200pt]  {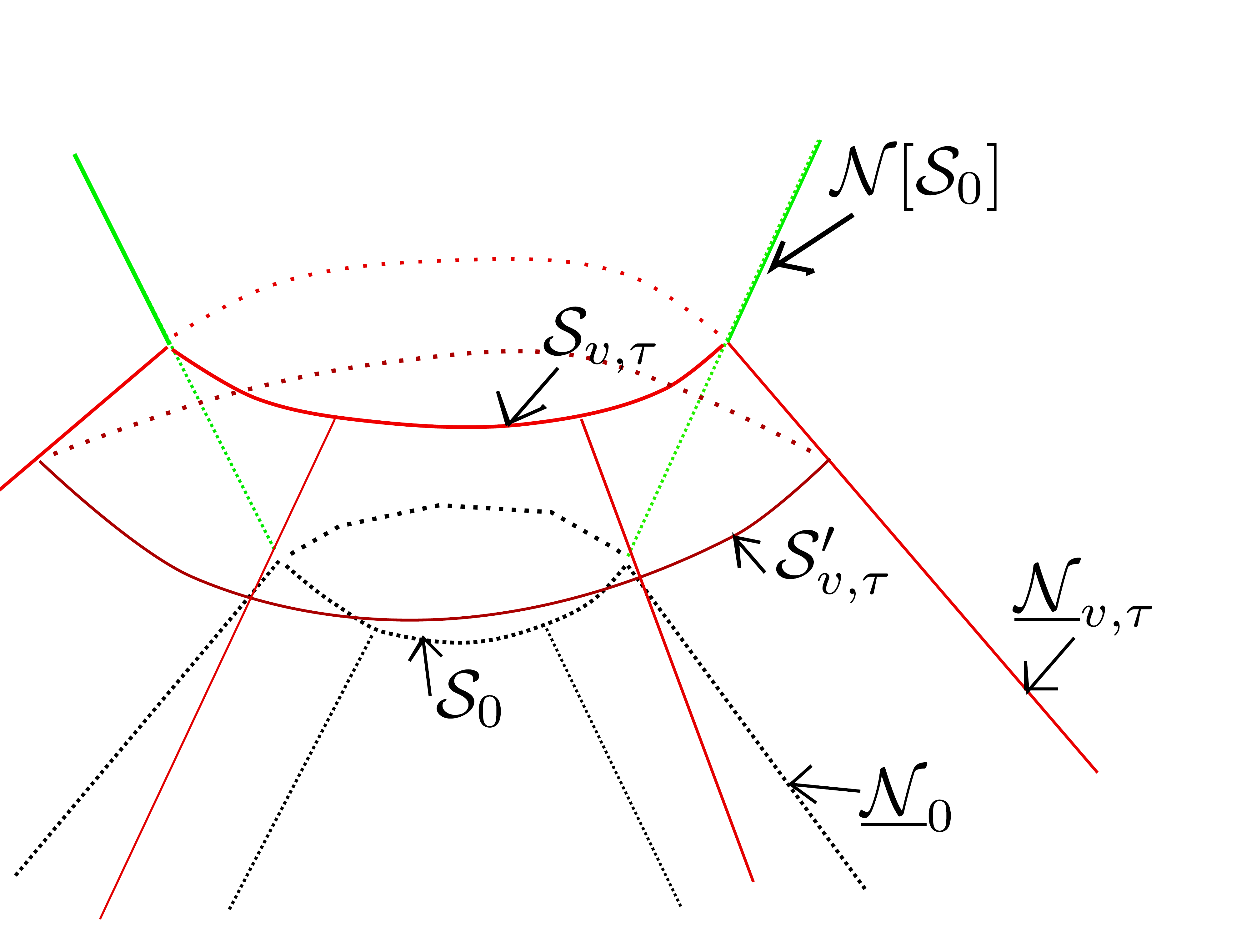} 
\caption{The ``old'' null surface $\lmcN_0$, and the ``new'' $\lmcN_{v,\tau}$ 
emanating from $\mc{S}^\prime_{v,\tau}$.}
\end{figure}

For future reference, we let:
\begin{equation}
\label{varpi}
\begin{split}
&\ol{\mc{L}}(e^v):=\tau^{-1}[tr\ul{\chi}_{v,\tau}-2]
\\&=[-2\rho[\mc{S}_0]+\hat{\chi}_{ab}[\mc{S}_0]
\cdot\hat{\ul{\chi}}^{ab}[\mc{S}_0]\\&-2div\zeta[\mc{S}_0] -2|\zeta[\mc{S}_0]|^2]e^v-2   
\Delta_{\mc{S}_0} e^v - 4  (  \zeta^a[\mc{S}] \nabla_ae^v)
\\&+ e^v |\hat{\chi}|^2[\mc{S}_0][
-2\rho[\mc{S}_0]-2div\zeta [\mc{S}_0]-2|\zeta[\mc{S}_0]|^2+\hat{\chi}
[\mc{S}_0]\hat{\ul{\chi}}[\mc{S}_0]]^{-1}\big{(}-2
-|\hat{\ul{\chi}}|^2[\mc{S}_0]\big{)}. 
\end{split}
\end{equation} 
Note that the operator $\ol{\mc{L}}$ can also be expressed in the form: 
\beq
\label{1st.op}
\ol{\mc{L}}(e^v)=[\Delta_{\mc{S}_0}+O^\delta_2(1) \partial_i +(\frac{1}{2m^2}+O^\delta_2(1))]e^v.
\eeq
(The use of the symbol $O^\delta_2(1)$ is an abuse of notation, since the 
functions in question do not depend on $\la$--here it merely means that the functions involved as
 well as their first and second rotational derivatives are bounded by $\delta$).

We remark also that the area of $ \mc{S}'_{v,\tau}$  is not lesser than $\mc{S}_0$: 
\begin{lemma}
\label{areas}
With $\mc{S}'_{v,\tau}$ as above (see Figure 2), 
$${\rm Area}[\mc{S}'_{v,\tau}]\ge  {\rm Area}[\mc{S}_0].$$
Furthermore we have equality in the above if and only if $tr\chi[\tilde{\mc{S}}]=0$ 
for each sphere $\tilde{\mc{S}}\subset\mc{N}[\mc{S}_0]$ contained 
between $\mc{S}_0$ and 
$\mc{S}_{v,\tau}$ on $\mc{N}[\mc{S}_0]$, and moreover
$F(v,\tau)=0$. 
\end{lemma}
 
 {\it Proof of Lemma \ref{newMOTS}:} We start by invoking formula 
 \eqref{eq.cf_chibar.trace} to find that 
 \beq
 \begin{split}
& tr\ul{\chi}[\mc{S}_{v,\tau}]=tr\ul{\chi}[\mc{S}_0]+\tau \{[-2\rho[\mc{S}]+\hat{\chi}_{ab}[\mc{S}_0]
\cdot\hat{\ul{\chi}}^{ab}[\mc{S}_0]\\&-2div\zeta[\mc{S}_0] -2|\zeta[\mc{S}_0]|^2]e^v-2   
\Delta_{\mc{S}_0} e^v - 4  (  \zeta^a[\mc{S}_0] \nabla_ae^v)\}+o_2(\tau)
\end{split}
 \eeq
 On the other hand, letting $\chi$ stand for the second fundamental form on 
 $\mc{N}[\mc{S}_0]$, the first 
 formula in \eqref{eq.structure_ev} tells us that 
 \beq
 tr\chi[\mc{S}_{v,\tau}]=-\tau e^v|\hat{\chi}|^2[\mc{S}_0]+o_2(\tau)
 \eeq
 Now, using the mean-value theorem,
  we find that there exists a function $F(v,\tau)$ 
 so that $\mc{S}^\prime_{v,\tau}$ (as in defined in the theorem statement)
 is marginally outer trapped, and moreover: 
 
 \beq
 \label{Fdef}
F(v,\tau)=-\frac{tr\chi^L[\mc{S}_{v,\tau}]}{\nabla_{\ul{L}_{v,\tau}} 
tr\chi^L[\mc{S}_{v,\tau}]}+o_2(\tau)=\frac{\tau e^v 
|\hat{\chi}|^2[\mc{S}_0]}{\nabla_{\ul{L}_{v,\tau}} tr\chi[\mc{S}_{v,\tau}]}+o_2(\tau). 
\eeq 
To show (\ref{new.trulchi}),   we first recall \eqref{var.trchi}. 

\beq
\begin{split}
\label{deriv}
&\nabla_{\ul{L}_{v,\tau}} tr\chi[\mc{S}_{v,\tau}]=-\frac{1}{2}tr\chi [\mc{S}_{v,\tau}] tr\ul{\chi}[\mc{S}_{v,\tau}]-2\rho[\mc{S}_{v,\tau}]-
2div\zeta [\mc{S}_{v,\tau}]-2|\zeta[\mc{S}_{v,\tau}]|^2
\\&-\hat{\chi}
[\mc{S}_{v,\tau}]\hat{\ul{\chi}}[\mc{S}_{v,\tau}]
\end{split}
\eeq
Note in particular that the assumed closeness to the Schwarzschild space-time implies that
$\nabla_{\ul{L}_{v,\tau}} tr\chi[\mc{S}_{v,\tau}]$ 
is
 bounded below by $4^{-1}m^{-2}$,
 for $\tau\in [0,1)$. 
Using \eqref{var.trchi} we derive that: 

\beq
\label{formula.F}
\begin{split}
&F(v,\tau)=    \tau e^v|\hat{\chi}|^2[\mc{S}_0]\{
-2\rho[\mc{S}_0]-2div\zeta [\mc{S}_0]-2|\zeta[\mc{S}_0]|^2-\hat{\chi}
[\mc{S}_0]\hat{\ul{\chi}}[\mc{S}_0]\}^{-1}
+o_2(\tau).
\end{split}
\eeq
Therefore (\ref{new.trulchi}) follows by invoking the first (traced) formula 
in \eqref{eq.structure_ev'} to obtain:

\beq
\begin{split}
\label{New-expn}
&tr\ul{\chi}[\mc{S}'_{v,\tau}] =tr\ul{\chi}[\mc{S}_{v,\tau}]+F(v,\tau)[-\frac{1}{2}
(tr\ul{\chi})^2[\mc{S}_{v,\tau}]
-|\hat{\ul{\chi}}|^2[\mc{S}_{v,\tau}]]+o_2(\tau)
\\&=tr\ul{\chi}[\mc{S}_{v,\tau}]
+F(v,\tau)[-\frac{1}{2}2^2
-|\hat{\ul{\chi}}|^2[\mc{S}_0]\big{)}+o_2(\tau)
\\&=tr\ul{\chi}[\mc{S}_0]+\tau \{[-2\rho[\mc{S}_0]-\hat{\chi}_{ab}[\mc{S}_0]
\cdot\hat{\ul{\chi}}^{ab}[\mc{S}_0]\\&-2div\zeta[\mc{S}_0] -2|\zeta[\mc{S}_0]|^2]e^v-2   
\Delta_{\mc{S}_0} e^v - 4  (  \zeta^a[\mc{S}_0] \nabla_ae^v)\}
\\&-\tau e^v  |\hat{\chi}|^2[\mc{S}_0] \{
-2\rho[\mc{S}_0]-2div\zeta [\mc{S}_0]-2|\zeta[\mc{S}_0]|^2-\hat{\chi}
[\mc{S}_0]\hat{\ul{\chi}}[\mc{S}_0]\}^{-1}\big{(}2
+|\hat{\ul{\chi}}|^2[\mc{S}_0]\big{)}+o_2(\tau)
\end{split}
\eeq 
This proves \eqref{new.trulchi}.$\Box$
\newline

We remark that  formula \eqref{New-expn} can be used to define an appropriate {\it affine}
parameter on $\lmcN_{v,\tau}$: Indeed, 
the above shows that there exists a function $A(v,\tau)$

\beq
\begin{split}
\label{right.scale}
&A(v,\tau)= -\tau \{[-2\rho[\mc{S}_0]-\hat{\chi}_{ab}[\mc{S}_0]
\cdot\hat{\ul{\chi}}^{ab}[\mc{S}_0]\\&-2div\zeta[\mc{S}] -2|\zeta[\mc{S}_0]|^2]e^v-2   
\Delta_{\mc{S}_0} e^v - 4  (  \zeta^a[\mc{S}_0] \nabla_ae^v)\}
\\&-\tau e^v  |\hat{\chi}|^2[\mc{S}_0] \{
2\rho[\mc{S}_0]-2div\zeta [\mc{S}_0]-2|\zeta[\mc{S}_0]|^2-\hat{\chi}
[\mc{S}_0]\hat{\ul{\chi}}[\mc{S}_0]\}^{-1}\big{(}2
+|\hat{\ul{\chi}}|^2[\mc{S}_0]\big{)}+o_2(\tau)
\end{split}
\eeq
so that if we let
\beq
\label{change}
\ul{L}^\sharp_{v,\tau,\nu }:=(1+ A(v,\tau))
\ul{L}_{v,\tau},
\eeq
and denote by $tr\ul{\chi}^\sharp_{v,\tau}$ the corresponding null second fundamental form of $\lmcN_{v,\tau}$ then:

 \beq
 tr\ul{\chi}^\sharp_{v,\tau}[\mc{S}'_{v,\tau}]=2.
 \eeq

Note for future reference that $A(v,\tau)$ can be expressed in the form: 
\beq
\label{right.scale'}
A(v,\tau)=\tau\{2\Delta_{\mc{S}_0} +O^\delta_2(1)\cdot \nabla -[\frac{1}{2m^2}
+O^\delta_2(1)]\}e^v +o_2(\tau),
\eeq
and in particular we can re-express \eqref{change} as:

\beq
\label{right.scale2}
\ul{L}^\sharp_{v,\tau,\nu }=e^{\tau \{2\Delta_{\mc{S}_0} +O^\delta_2(1)\cdot \nabla -[\frac{1}{2m^2}
+O^\delta_2(1)]\}e^v+o_2(\tau)}.
\eeq

 {\it Proof of Lemma \ref{areas}:} It suffices to show a localized version of our claim. 
 Consider a natural map between $\mc{S}_0$ and $\mc{S}'_{v,\tau}$
 which identifies the  points on  null generators
  of $\mc{N}[\mc{S}_0]$, $\lmcN_{v,\tau}$ that intersect on $\mc{S}_{v,\tau}$ 
  (see the East and West ``poles'' on $\mc{S}_{v,\tau}$ in Figure 2.
    We can thus identify area 
  elements on these spheres; thus for a given triplet of 
  points $P_1,P_2\in\mc{N}[\mc{S}_0]$ and $P_2, P_3\in\lmcN_{v,\tau}$
  that lie on the same null generators of $\mc{N}[\mc{S}_0]$, $\lmcN_{v,\tau}$
  respectively, we think of $dV_{\mc{S}_{v,\tau}}(P_1)(dV_{\mc{S}_0}(P_2))^{-1}$, 
  $dV_{\mc{S}_{v,\tau}}(P_3)(dV_{\mc{S}'_{v,\tau}}(P_2))^{-1}$
   as a numbers.
Letting $x$ stand for a point in the null segment $(P_1P_2)$ we 
then derive: 
  
\beq
log[dV_{\mc{S}_{v,\tau}}(P_2)(dV_{\mc{S}_0}(P_1))^{-1}]=\int_0^{\tau e^v} tr\chi^L(x)dx
\eeq 
In particular since (by the Raychaudhuri equation) 
$tr\chi^L$ is a non-increasing, non-positive function along $(P_1P_2)$, 
we derive: 
\beq
\label{estim1}
|log[dV_{\mc{S}_{v,\tau}}(P_1)(dV_{\mc{S}_0}(P_2))^{-1}]|\le \tau e^v |tr\chi^L(P_2)|
\eeq 
We now use $tr\ul{\chi}^{\ul{L}_{v,\tau}}$ on $\lmcN_{v,\tau}$
 to derive (letting $x\in (P_2P_3)$): 
\beq
\begin{split}
\label{estim2}
&log[dV_{dV_{\mc{S}'_{v,\tau}}(P_3)(\mc{S}_{v,\tau}}(P_2))^{-1}]=
\int_0^{F_{P_2}(v,\tau)} tr\ul{\chi}^{\ul{L}_{v,\tau}}(x)dx
\\&\ge \frac{1}{2}F_{P_2}(v,\tau)tr\ul{\chi}^{\ul{L}_{v,\tau}}(P_2)\ge 
\frac{tr\chi^L(\mc{S}_{v,\tau})}{4m^2},
\end{split}
\eeq
using \eqref{formula.F} and \eqref{connectionS0} in the second inequality.

Thus, adding (\ref{estim1}) and (\ref{estim2})  we derive
(for $\tau$ small enough)  the inequality:
\beq
dV_{\mc{S}'_{v,\tau}}(P_3)(dV_{\mc{S}_0}(P_1))^{-1}\ge 1.
\eeq
Moreover, clearly we have equality if and only if $tr\chi^L(P_2)=0$. In this case, the Raychaudhuri equation 
implies that $tr\chi^L(P)=0$ for all $P\in (P_1P_2)$; moreover clearly $F_{P_2}(v,\tau)=0$. 

Thus, we have derived that: 
\beq
dV_{\mc{S}'_{v,\tau}}\ge dV_{\mc{S}_0}
\eeq
with equality if and only if $tr\chi^L=0$ in $\mc{N}[\mc{S}_0]$ 
between $\mc{S}_0$ and $\mc{S}_{v,\tau}$ and $F_{v,\tau}=0$ on $\mc{S}_{v,\tau}$. $\Box$
\newline

 We note for future reference that by a similar argument (integrating the structure equations 
 \eqref{eq.structure_ev}, \eqref{eq.structure_ev'} in the $L$ and $\ul{L}$-directions 
 respectively, and using \eqref{connectionS0}) shows that  
 for all $v,\tau$ as in Lemma \ref{newMOTS} we have 
 
\beq
\label{future.ref}
|\partial^{(i)}\zeta|_{\mc{C}^0(\mc{S}'_{v,\tau})}\le\delta, 
|\partial^{(i)}\chi|_{\mc{C}^0(\mc{S}'_{v,\tau})}\le\delta
\eeq
 for all $i\le 2$.
\newline

 The main method towards proving our result will be by expoliting a monotonicity property 
 enjoyed by the Hawking mass on null hypersurfaces 
that are perturbations of the shear-free null hypersurfaces in Schwarzschild. We review this in
 the next subsection.

 \section{Monotonicity of Hawking mass on  smooth null surfaces.} 
\label{sec-monotonicity}

In this section we will review certain well-known monotonicity properties of the Hawking mass 
which go back to \cite{H} (see also \cite{Sauter}, whose notation we largely follow).
We begin by understanding the asymptotic behaviour of the relevant Ricci coefficients for all 
hypersurfaces $\lmcN_{v,\tau}$ 
that we consider; we then proceed to study the behaviour of these coefficients under 
perturbations of the underlying hypersurfaces.

\subsection{The asymptotics of the expansion of null surfaces  relative to an affine 
vector field. } 
 
 Understanding the asymptotic behaviour  of the expansion of a smooth 
 past-directed outgoing $\lmcN$ 
will be necessary for the construction and study of our luminosity foliation.

 We introduce a convention: When we write a tensor (defined over level spheres in $\lmcN$) 
with lower case indices $a,b$ we think of it as an abstract tensor field. 
When we use upper-case indices, we will be referring to its components evaluated against the vector fields $e^A, A=1,2$ defined in Definition \ref{frame-constr}.
  
 \begin{lemma}
 \label{trulchi}
 Consider any infinite smooth past-directed outgoing null surface 
 $\lmcN$. Assume that $\la$ is an affine parameter on $\lmcN$ and let $\ul{L}^\la$ be the corresponding affine vector field. 
 Assume that $\ul{\alpha}_{ab}(\la)$ defined relative to $\ul{L}^\la$ satisfies the fall-off 
 condition
\[
\ul{\alpha}_{AB}(\la)\in O^\delta_2(\la^{-3-\epsilon}).
\]
 Assume also that $|tr\ul{\chi}(x)-2|$ and $|\hat{\ul{\chi}}|(x)$
are sufficiently small on $\mc{S}=\partial\lmcN$. 
Then  there exist 
 continuous and bounded  functions (tensors) $a(\phi,\theta,\lambda), h_{ab}(\phi,\theta,\lambda)
 \in O^\delta(1)$ 
which converge to continuous  limits $a(\phi,\theta), b_{ab}(\phi,\theta)$ over 
$\mc{S}_\infty:=\partial_\infty \lmcN$  as 
$\lambda\to\infty$  so that:
 
 \beq
 \label{asymptotics1}
 tr\ul{\chi}^\lambda=\frac{2}{\lambda}+\frac{a(x,\la)}{\lambda^2} , \hat{\ul{\chi}}_{ab}=\frac{h_{ab}(x,\la)}{\lambda^2}. 
 \eeq
 Furthermore the first and second spherical derivatives $\partial_{\phi^i}$, 
 $\partial^2_{\phi^i\phi^j}$
  of the functions $a(x,\la), h_{ab}(x,\la)$ remain in $O^\delta(1)$ 
   and have continuous limits as 
  $\la\to\infty$. 
  \end{lemma}

  We also note for future reference that Lemma \ref{trulchi} can also be applied 
  to \emph{families} of infinite 
  null surfaces: 
  Consider any 1-parameter family of infinite smooth outgoing 
  null surfaces $\lmcN_\tau$, $\tau\in [0,1]$, let $\la$ is an affine parameter on each 
  of these,  and assume that the corresponding Weyl curvature component 
  $\ul{\alpha}_\tau$ 
  satisfies the same fall-off condition 
\[
|\ul{\alpha}_\tau|\in O^\delta_2(\la^{-3-\epsilon}).
\]
  Then the bounds \eqref{asymptotics1} hold for $tr\ul{\chi}^\la_\tau$, 
  $(\hat{\ul{\chi}}_{ab})_\tau$.


  
  
We also note a consequence of Lemma \ref{trulchi}: If we let 
$\nasla_A$ be the (intrinsic) connection on the level spheres $\mc{S}_\la$
of the affine parameter on $\lmcN$, then for $A,B\in \{1,2\}$:
\beq
\label{decay-conn}
\nasla_Ae^B\in O_2(\la^{-1}).
\eeq

We wish to study the first variation of the quantities above in $\tau$, for the surfaces 
$\lmcN_{v,\tau}$ that we consider. To do this, we need the following definition:
  
\begin{definition}
We say that a 1-parameter family of smooth outgoing past-directed infinite null 
surfaces $\lmcN_\tau$ as above is 
of class $\mc{R}(\delta)$ if 
there exists a natural map: 
$\Phi_\tau:\lmcN\to\lmcN_\tau$ with $\Phi_\tau(\la,\phi,\theta)=(\la,\phi_\tau,\theta_\tau)$ so that:

\begin{itemize} 
\item the componets of $\ul{\alpha}_\tau$ 
measured relative to the frame $\la^{-1}(\Phi_\tau)_*\partial_\phi, \la^{-1}(\Phi_\tau)_*\partial_\theta$ 
are differentiable in $\tau$ and obey the bound 
\[
|\partial_\tau(\ul{\alpha}_{AB})_\tau|\in O^\delta_2(\la^{-3-\epsilon}).
\]

 \item The metric components $(\mind_{ab})_{\tau}(\la)$ measured relative 
 to the same frame satisfy: $\partial_\tau (\gamma_{\tau})_{AB}(\la)=O(1)$.
\item{$tr\ul{\chi}_\tau=2$ at $\mc{S}_\tau:=\partial\lmcN_\tau$ and 
$|(\hat{\ul{\chi}}_{AB})_\tau|_{\mc{S}_\tau}\in O^\delta_2(\la^{-2})$.}
\end{itemize}
\end{definition}

  \begin{lemma}
  \label{trulchi.var}
Consider a 1-parameter family of infinite null surfaces $\lmcN_{v,\tau}$, of 
class $\mc{R}(\delta)$ as described above. Assume that there exist functions 
$\mc{F}(v), \mc{G}(v)\in\mc{C}^2(\phi,\theta)$ 
 (both depending only on $\phi,\theta$, and the function $v(\phi,\theta)$)
and tensors $f^{ab}\in O_2^1(1)$ ,$\ol{f}_{ab}\in O_2^\delta(\la^{-3-\epsilon})$ 
depending on 
$\phi,\theta,\la$
so that the first variation of the metric and curvature components is given by:
\beq
\label{eq.trulchi.var}
\dot{\mind}^{AB}_v=f^{AB}(\la)\mc{F}(v), (\dot{\ul{\alpha}}_{AB})_v=\ol{f}_{AB}(\la)
\mc{G}(v).
\eeq

  Then the function $\dot{a}_v:=\partial_\tau|_{\tau=0}a_{v,\tau}$, 
  and the components of the tensor field
  $(\dot{h}_{ab})_v:=\partial_\tau|_{\tau=0} (h_{ab})_{v,\tau}$ 
can be expressed in the form

\beq
\dot{a}_v(\la)= b_1(\la)\mc{F}[v]+b_2(\la)\mc{G}[v], (\dot{h}_{AB})_v(\la)= b_3(\la)\mc{F}[v]+b_4(\la)\mc{G}[v], 
\eeq  
  where $b_i\in O^\delta_2(1)$, $i=1,2,3,4$.
  

 \end{lemma}

 {\it Proof of Lemma \ref{trulchi}:}
 We refer to the evolution equation \eqref{eq.structure_ev'} for $\ul{\chi}$. 
 This is a tensorial equation, while 
 we are interested in the components of  the tensors $\ul{\chi}$. To do this, we first
 recall the (tensorial) evolution  equations on $\tr\ul{\chi}$ and $\hat{\ul{\chi}}$
 on  a null surface, see the first equation in 
 \eqref{eq.structure_ev'}: 
 
 \beq
 \label{2forms}
 \nasla_\la \tr\ul{\chi}=-\frac{1}{2}(\tr\ul{\chi})^2-|\hat{\ul{\chi}}|^2, \nasla_\la \hat{\ul{\chi}}_{ab}=-\tr\ul{\chi}\hat{\ul{\chi}}_{ab}-\ul{\alpha}_{ab}.
 \eeq
 
As discussed above, we let $(\hat{\ul{\chi}}_{AB})$ 
  be the evaluation $\hat{\ul{\chi}}(e_A, e_B)$; accordingly we let $(\ul{\alpha}_{AB})$. (The $e_A, e_B$ are among $\Phi,\Theta$). 
  Using the fact  that $\ul{L}$ commutes with the two vector fields $\partial_\phi$ and 
  $\partial_\theta$, 
we derive the formulas: 
 \beq
 \begin{split}
&\nabla_{\ul{L}}\ul{L}=0, g(\nabla_{\ul{L}}e_A, e_B)=-\frac{1}{\la}\mind_{AB}
+\ul{\chi}_{AB}, 
g(\nabla_{\ul{L}}e_A,\ul{L})=-g(\nabla_{\ul{L}}\ul{L}, e_A)=0, 
\\& 
g(\nabla_{\ul{L}}e_A, L)=\zeta_A.
\end{split}
\eeq
Also
\beq\begin{split}
&g(\nabla_{\ul{L}}L,L)=0, 
g(\nabla_{\ul{L}}L,\ul{L})=-g(\nabla_{\ul{L}}\ul{L},L)=0,
\\& g(\nabla_{\ul{L}}L, e_A)=-g(\nabla_{\ul{L}}e_A,L)=
-g(\nabla_{e^A}\ul{L},L)=-2\zeta_A.
\end{split}
\eeq

Furthermore, recall that $\ul{\chi}_{AB}$ is a symmetric $(0,2)$ tensor field over the level sets of $\la$.  
We let $(\ul{\chi}^\sharp)_A^B$ be the corresponding $(1,1)$-tensor defined via the relation: 

\[
(\ul{\chi}^\sharp)_A^Be_B=\ul{\chi}(e_A, e_B).
\]

Therefore:
\beq
\label{frame.evol}
\nabla_{\ul{L}}\ul{L}=0, \nabla_{\ul{L}}e_A=-\frac{1}{\la}e_A
+(\ul{\chi}_A^\sharp)^Be_B+\zeta_A \ul{L}, 
\nabla_{\ul{L}}L=\zeta^Ae_A.
\eeq
Then  by the definition \eqref{nasla.def} 
of $\nasla_{\ul{L}}$ and from \eqref{frame.evol} we have:
 \[
 \nasla_{\ul{L}}e_A=-\frac{1}{\la}e_A+(\ul{\chi}^\sharp)^B_Ae_B
 \]

We also recall the definitions of $\chi_{AB}$, and note that:
\[
g(\nabla_AL, L)=0, g(\nabla_AL,\ul{L})=-g(\nabla_A\ul{L},L)=-2\zeta_A.
\]
In view of this we find:

\beq
\label{nablaL}
\nabla_AL=\chi_A^Be_B-\zeta_AL.
\eeq
We derive 
\beq
\label{metric.evol}
\begin{split}
&\ul{L}\mind_{AB}=\ul{L}g(e_A,e_B)=g(\nabla_{\ul{L}}e_A, e_B)+
g(e_A, \nabla_{\ul{L}}e_B)=-\frac{2}{\la}\mind_{AB}+2\ul{\chi}(e_A, e_B)
\\&=
-\frac{2}{\la}\mind_{AB}+\mind_{AB}\tr\ul{\chi}+2\hat{\ul{\chi}}_{AB}=
\frac{a}{\la^2}\mind_{AB}+2\la^{-2}h_{AB}. 
\end{split}
\eeq

 Then, using the second formula in \eqref{2forms} and the definition of covariant differentiation, we find, for $e_A, e_B$ among $\Phi,\Theta$
 
 \[
 \begin{split}
 \label{chi-bar-hat}
& \nasla_{\la}\big{(}  \hat{\ul{\chi}}_{AB} \big{)}=\big{(} \nasla_\la\hat{\ul{\chi}}  \big{)}(e_A, e_B)+\hat{\ul{\chi}}(\nasla_\la e_A,e_B)+
 \hat{\ul{\chi}}(\nasla_\la e_B,e_A)
\\&=-\frac{2}{\la}\hat{\ul{\chi}}(e_A, e_B) -\tr\ul{\chi}\hat{\ul{\chi}}(e_A, e_B)-\ul{\alpha}_{AB}+\ul{\chi}_{AC}\hat{\ul{\chi}}^C_B+\ul{\chi}_B^C
\hat{\ul{\chi}}_{CA}
 \\&=-\frac{2}{\la}\hat{\ul{\chi}}(e_A, e_B)+2\hat{\ul{\chi}}_{AC}
 \hat{\ul{\chi}}^C_B-\ul{\alpha}_{AB}
\end{split} 
 \]
 
 Now, let 
 $a:=\lambda^2(tr\ul{\chi}^\lambda-\frac{2}{\lambda}), 
 h_{ab}:=\lambda^2\hat{\ul{\chi}}_{ab}$.

 Then using the  \eqref{2forms} 
  we derive the following equations for $a, h$:
 \beq
 \label{1st.eq}
 \frac{d}{d\lambda}a=-\frac{1}{\lambda^2}a^2-\frac{h_{AB}h_{CD}\mind^{AC}\mind^{BD}}{\lambda^2}.
 \eeq
 
 \beq
 \label{2nd.eq}
 \frac{d}{d\lambda} (h_{AB})+\frac{2}{\la^2}h_{AD}\cdot h_{CB}\mind^{CD}=
- \lambda^2(\ul{\alpha}_{AB}).
 \eeq

Now, combining the above two equations with \eqref{metric.evol}, the claimed asymptotic 
behavior along with the bound\footnote{This follows from the definition of $\ul{\chi}$ and the asymptotics in \eqref{asymptotics1}.} 
\[
\mind_{AB}[\mc{S}_\la]=\mind_{AB}[\mc{S}_0]+O^\delta(1)
\]
 follow by a simple bootstrap argument, using  
\[
\la^2\ul{\alpha}_{AB}\in O^\delta(\la^{-1-\epsilon}).
\]

To derive the claim on the spherical derivatives of $a,h$, we take first one 
derivative $\partial_{\phi^i}$ 
of (\ref{1st.eq}), (\ref{2nd.eq}), \eqref{metric.evol}. 
We then obtain  a system of three linear first
 order ODEs:
 \beq
 \label{1st.eq'}
 \begin{split}
& \frac{d}{d\lambda} (\partial_{\phi^i}a)=-\frac{2}{\lambda^2}
 (\partial_{\phi^i}a) a-\frac{2}{\la^2}
 \sum_{A,B, C, D=1}^2[(\partial_{\phi^i}h_{AB})h_{CD}
 \mind^{AC}\mind^{BD}
 \\&+(\partial_{\phi^i}\mind^{AC})
 h_{AB}h_{CD}\mind^{BD}]
 \\&=O^\delta(\la^{-2})(\partial_{\phi^i}a)+ \sum_{C,D=1}^2
 O^\delta(\la^{-2})
 (\partial_{\phi^i}h_{CD})+\sum_{A,C=1}^2 O^\delta(\la^{-2})
 (\partial_{\phi^i}\mind^{AC}),
\end{split}
 \eeq

 \beq
 \label{2nd.eq'}
\begin{split}
& \frac{d}{d\lambda} (\partial_{\phi^i}{h}_{AB})=
-\frac{4}{\la^2}(\partial_{\phi^i}{h}_{AC})h_{BD}\mind^{CD}
-\frac{2}{\la^2}(\partial_{\phi^i}\mind^{CD})h_{AC}h_{BD}
+  \lambda^2(\partial_{\phi^i}{\ul{\alpha}}_{AB})
 \\&\sum_{C,D=1}^2O^\delta(\la^{-2})(\partial_{\phi^i}{h}_{CD})+
 \sum_{C,D=1}^2O^\delta(\la^{-2})(\partial_{\phi^i}h_{CD})-
 \la^2(\partial_{\phi^i}\ul{\alpha}_{AB}),
\end{split} 
 \eeq

\beq
\label{3rd.eq'}
\begin{split}
&\frac{d}{d\la}(\partial_{\phi^i}\mind_{AB})=\frac{\partial_{\phi^i}a}{\la^2}\mind_{AB}+
\frac{a}{\la^2}\partial_{\phi^i}\mind_{AB}+2\la^{-2}\partial_{\phi^i}\mind_{AB}
\\&O(\la^{-2})\partial_{\phi^i}a+O(\la^{-2})\partial_{\phi^i}\mind_{AB}.
\end{split}
\eeq 
 
 The  $O(\la^{-2})$ terms in the last lines of the above equations  follow from 
 the bounds we have already derived in the previous step. 
 
Our claim on the first derivatives thus follows from standard formulas for this first order system 
of linear equations.  
 Then replacing this in \eqref{1st.eq}, we
 derive the claim for $\partial_{\phi^i}a$. The claim for the second
  derivatives follows by taking a further rotational derivative of the above equations 
  and repeating the same argument. 
 $\Box$ 
 
 For future reference, we note that the above 
imply the  bounds 
\beq
\label{gamma.bounds} 
\mind_{AB}[\mc{S}_\la]=\mind_{AB}[\mc{S}_0]+O^\delta_2(1),
\eeq
which also capture  up to two of 
 the rotational derivatives of $\mind_{AB}$. 
\newline

 {\it Proof Lemma \ref{trulchi.var}:} We again 
 consider the evolution equations 
(\ref{1st.eq}), \eqref{2nd.eq}, \eqref{metric.evol} for  $a_{v,\tau}$ and the evaluation of $h_{v,\tau}$ against 
frame $e^1, e^2$. (So now $a, h_{AB}$ depend on the parameters $v,\tau$).

We then consider the $\partial_\tau|_{\tau=0}$-derivative of this
system. (Recall that $\partial_\tau|_{\tau=0}$ stands for a Jacobi field $J$--see \eqref{Jac}).
Since $\la_{v,\tau}$ and $\partial_\tau$ commute by construction, we
 find:

 \beq
 \label{1st.eq''}
 \begin{split}
& \frac{d}{d\lambda} (\dot{a}_{v})=-\frac{2}{\lambda^2}
 (\dot{a}_{v}) a-\frac{2}{\la^2}
 \sum_{A,B, C, D=1}^2[(\dot{h}_{AB})_vh_{CD}\mind^{AC}
 \mind^{BD}+(\dot{\mind}^{AC})
 h_{AB}h_{CD}\mind_{v,\tau}^{BD}]
 \\&=O_2^\delta(\la^{-2})(\dot{a}_{v})+ O_2^\delta(\la^{-2})
 (\dot{h}_{ij})_{v}+\sum_{A,C=1}^2 O_2^\delta(\la^{-2})
 (\dot{\mind}_{v}^{AC})
\end{split}
 \eeq

 \beq
 \label{2nd.eq''}
\begin{split}
& \frac{d}{d\lambda} (\dot{h}_{AB})_{v}=
-\frac{4}{\la^2}(\dot{h}_{AC})_vh_{BD}\mind^{CD}
-\frac{2}{\la^2}(\dot{\mind}^{CD})h_{AC}h_{BD}
-  \lambda^2(\dot{\ul{\alpha}}_{AB})_{v}
 \\&=\sum_{C,D=1}^2O_2^\delta(\la^{-2})(\dot{h}_{CD})_{v}+
 \sum_{C,D=1}^2O_2^\delta(\la^{-2})(\dot{h}_{CD})_{v}-
 \la^2(\dot{\ul{\alpha}}_{AB})_{v}
\end{split} 
 \eeq
\beq
\label{3rd.eq''}
\begin{split}
&\frac{d}{d\la}(\dot{\mind}_{AB})=\frac{\dot{a}}{\la^2}\mind_{AB}+
\frac{a}{\la^2}\dot{\mind}_{AB}+2\la^{-2}\dot{\mind}_{AB}
\\&O^\delta_2(\la^{-2})\dot{a}+O^\delta_2(\la^{-2})\dot{\mind}_{AB}.
\end{split}
\eeq 

Thus, recalling that $\dot{a}_v=0$ at $\la=1$,
 our result again follows by the above first-order system of linear ODEs, by
integration of these first order equations.   
This completes the proof of our Lemma.$\Box$
\newline

We note that \eqref{asymptotics1} together with the formula\footnote{The indices $a,b$ 
below take values among the coordinate vector fields $\partial_{\phi^1}, \partial_{\phi^2}$.}
\beq
\label{pre.two.gs}
\mind_{ab}(\{\la=r_1(\phi,\theta)\})-\mind_{ab}(\{\la=r_2(\phi,\theta)\})=
2\int_{\{\la=r_1(\phi,\theta)\}}^{\{\la=r_2(\phi,\theta)\}}\ul{\chi}^{\ul{L}}_{ab}(\la),
\eeq
where $\la$ is any chosen affine parameter on $\lmcN_{v,\tau}$ 
and $\ul{L}$ the associated affine 
vector field, implies the existence of the limit (where $a,b$ are evaluated against the coordinate vector field $\partial_{\phi^1}, \partial_{\phi^2}$)
\beq
\label{metric.infty}
(\gamma^{\infty,\la}_{v,\tau})_{ab}:=lim_{\la_{v,\tau}\to\infty} \la_{v,\tau}^{-2}
\mind_{ab}[\mc{S}_{\la_{v,\tau}}]
\eeq
on each $\lmcN_{v,\tau}$. 

\par Using \eqref{metric.infty}, applied to any affine function on each 
$\lmcN_{v,\tau}$ we 
naturally associate a ``metric at infinity'' (for the chosen affine function).
We also check that in view of the regularity in the angular directions
of $\ul{\chi}$ in \eqref{asymptotics1}, the limit of the Gauss curvatures of 
$\la^{-2}\mind_{ab}[\mc{S}_{\la_{v,\tau}}]$. This limit  in fact agrees with 
the Gauss curvature of the 
limiting metric $\gamma^{\infty,\la}_{v,\tau}$.
 As we will  see below, 
on any fixed $\lmcN_{v,\tau}$, any smooth change of the affine parameter $\la_{v,\tau}$ 
induces a conformal change on the metric at infinity, defined (for any affine parameter) 
via \eqref{metric.infty}.

We also note  a few useful facts  about the asymptotics of the Ricci coefficients
 $\zeta, \chi$ on the affinely parametrized null surfaces $\lmcN_{v,\tau}$:
 
 \begin{lemma}
 \label{zeta.chi}
 Given any surface $\lmcN_{v,\tau}$ in our space of perturbations, we claim that
   (letting $L_{v,\tau}$ be the null normal to the level sets 
 of $\la_{v,\tau}$ normalized so that $g(L_{v,\tau},\ul{L}_{v,\tau})=2)$:
 \beq
\label{zet.chi}
 \zeta^{\ul{L}_{v,\tau}}_A=O^\delta_1(\la_{v,\tau}^{-2}), \chi^{\ul{L}_{v,\tau}}_{AB}\in O^\delta_1(\la_{v,\tau}^{-1}).
 \eeq 
 and moreover: 
 \beq
 \label{lower.bd}
 tr\chi^{L_{v,\tau}}[\mc{S}_\la]\ge \frac{(\la_{v,\tau}-1)}{4m^2\la_{v,\tau}^2}.
 \eeq
 \end{lemma}
 
  {\it Proof:} Recall  from \eqref{future.ref} that 
\[
\chi^{L_{v,\tau}}[\mc{S}^\p_{v,\tau}]=0, \sum_{i=0}^2|\partial^{(i)}
\zeta_A^{L_{v,\tau}}[\mc{S}^\p_{v,\tau}]|\le 
\delta,  \sum_{i=0}^2|\partial^{(i)}\ul{\chi}_{AB}^{\ul{L}_{v,\tau}}|\le \delta
\]

  The claim on $\zeta$ follows from the second equation in 
  \eqref{eq.structure_ev'} (a linear ODE, given the bounds we have on $\ul{\chi}$) 
  by multiplying by $\la^2$ and evaluating against 
  $e_A$, $A=1,2$. 
  To derive the claim on the rotational derivatives of $\zeta_a$, we just differentiate 
  the evolution equations by $\partial_{\phi^i}$ and invoke the solution of first order ODEs, 
  along with the derived and assumed 
   bounds on the angular derivatives of $\ul{\chi}$ and $\ul{\beta}$. Once 
   the claim has been derived for $\zeta_a$, we refer to the third equation in 
   \eqref{eq.structure_ev'} and repeat the same
    argument for $\chi_{ab}$. This proves \eqref{zet.chi}.
    
    To derive \eqref{lower.bd} we invoke \eqref{asymptotics1} and multiply the equation 
    \eqref{var.trchi} by $\la$, and derive an equation:
  \beq
\label{evolution.chi}
    \frac{d}{d\la}[\la tr\chi]+\frac{atr\chi}{\la}=-2\la div\zeta -2\la |\zeta|^2-2\la\rho
+\la \hat{\chi}\hat{\ul{\chi}}.
    \eeq
    In the RHS of the above, all terms can be seen as perturbations of the main term 
    $-2\la\rho$, in view of the bounds we have already derived. Thus our result 
    follows by the bounds on $\rho$ in \eqref{curv.decay} and 
    treating the above as a first order ODE in $\la\tr\chi$, using  the smallness 
    of $\delta$ compared to $m$. 
$\Box$  
\newline

 For future reference we note two key facts: The first is that by integration of the evolution equation 
\eqref{evolution.chi}, we derive that $\la tr\chi$ has a limit over $\mc{S}^\infty_{v,\tau}$, which is a $\mc{C}^2$ function. 
We denote this limit by $tr\chi^{\la,\infty}$ to stress the dependence on the choice of affine function $\la$. ƒ
Note that by Lemmas \ref{trulchi}, \ref{zeta.chi}, and formula \eqref{GC},
  for any of the hypersurfaces $\lmcN=\lmcN_{v,\tau}$ this limit in fact agrees with the limit of the (renormalized) 
 Gauss curvatures of the level spheres $\mc{S}_\la$:

\beq
\label{key-note}
lim_{\la\to\infty} \la tr\chi^{\la}=2lim_{\la\to\infty} \la^2\mc{K}[\mc{S}_{\la}].
\eeq

In this connection, we make a note on 
the transformation law of $tr \chi$ on a given null surface $\lmcN$
which satisfies the conclusion of Lemmas \ref{trulchi} and \ref{zeta.chi}. We will be particularly interested in a function $\omega_\tau$ 
equal to the \emph{exponent} in  \eqref{right.scale2}. 
We let $\widetilde{tr\chi}^{\la,\infty}$ to be the limit of $\la tr\chi^\la[\mc{S}_\la]$ as $\la\to\infty$. 
We also let $\widetilde{tr\chi}^{\la^\p_\tau}$ to be the limit of $\la^\p_\tau tr\chi^\la[\mc{S}_{\la^\p_\tau}]$ as $\la^\p_\tau\to\infty$

Using that choice of $\omega$, the definition \eqref{metric.infty}
 and the asymptotics for $\chi,\ul{\chi}, \zeta$, 
we find that  \eqref{chi-trans} implies that letting $\la^\p_\tau$ be the new affine parameter defined via 
\eqref{two.las} for $\omega:=\omega_\tau$ we have:

\beq
\label{new.trchi}
\widetilde{tr\chi}^{\la^\p_\tau}=\widetilde{tr\chi}^\la+2[ \widetilde{tr\chi}^\la+
\Delta_{\gamma^{\infty,\la}}]\omega_\tau+o(\tau). 
\eeq

 \subsection{Monotonicity of Hawking Mass.}
 
In this subsection $\lmcN$ will stand for any infinite smooth past-directed 
outgoing null surface 
which satisfies the assumptions (and thus the conclusions) 
 of Lemma \ref{trulchi}.
 In particular recall that all the null surfaces 
$\lmcN_{v,\tau}$ considered in Lemma \ref{newMOTS} satisfy these assumptions.

We recall a fact essentially due to Hawking, \cite{H}: 
\begin{definition}
\label{luminosity}
Consider any foliation of $\lmcN$ by a smooth family of
 2-spheres $\mc{S}_s$, $s\in [1,+\infty)$. We consider the (unique) null geodesic generator
 $\ul{L}$ which is tangent to $\lmcN$ 
 and defined via: 
 \beq
\ul{L}s=1. 
 \eeq
We call $s$ a luminosity parameter if:
\beq
\label{lumin}
tr\ul{\chi}^{\ul{L}}[\mc{S}_s]=\frac{2}{s},
\eeq
We call the family $\mc{S}_s\subset\lmcN, s\in [1,\infty)$ a luminosity foliation 
of $\lmcN$. 
\end{definition}
In particular $tr\ul{\chi}^{\ul{L}}$ (defined relative to $\ul{L}$) is constant on each sphere 
$\mc{S}_s$.

A key property of luminosity foliations of $\lmcN$ is that the Hawking mass
is monotone for such a foliation, when $\lmcN$ is (extrinsically and intrinsically) close to 
the shear-free null surfaces in the Schwarzschild exteriors: 
\begin{lemma}
\label{monoton.mass}
We let $L$ be the conjugate null vector field 
to $\ul{L}$ on $\lmcN$ for the spheres $\mc{S}_s$ (i.~e.~$g(L,\ul{L})=2$, $L\perp \mc{S}_s, s\ge 1$) 
and also let $\ul{\chi}^{\ul{L}},\chi^L, \zeta^{\ul{L}}$ be the 
null expansions and torsion of the spheres $\mc{S}_s$ defined relative to $\ul{L},L$. 

For the luminosity foliation $\mc{S}_s$ of $\lmcN$ (satisfying (\ref{lumin})) we have: 
\beq
\label{monoton}
\frac{d}{ds}m_{\rm Hawk}[\mc{S}_s]=\frac{r[\mc{S}_s]}{32\pi}\int_{\mc{S}_s}tr\chi^L|\hat{\ul{\chi}}^{\ul{L}}|^2
+ tr\ul{\chi}^{\ul{L}}|\zeta^{\ul{L}}|^2dV_s. 
\eeq
\end{lemma}
In particular when $tr\chi^L[\mc{S}_s]\ge 0$ and $tr\ul{\chi}^{\ul{L}}[\mc{S}_s]\ge 0$
for all $s\ge 1$ (as will be the case for all $\lmcN_{v,\tau}$ that we consider here in view of
 \eqref{lower.bd}),  
$m_{\rm Hawk}[\mc{S}_s]$ is non-decreasing in $s$, in the outward direction.

{\it Proof:} The proof of this follows \cite{Sauter}, which elaborates the argument in
 \cite{H}.
We define a function $\kappa$ over $\lmcN$ defined via: 

\beq
\nabla_{\ul{L}}\ul{L}=\kappa\ul{L}.
\eeq
We then  recall the evolution equations on each $\mc{S}_s$:\footnote{We
 write $tr\ul{\chi},tr\chi, \zeta$ for short.}
  
   \beq 
   \begin{split}
& \nabla_{\ul{L}}tr\ul{\chi}^{\ul{L}}=-\frac{1}{2}(tr\ul{\chi})^2-\frac{1}{2}|\hat{\chi}|^2+\kappa tr\ul{\chi}, 
\\& \frac{d}{ds}dV_{\mc{S}_s}=tr\ul{\chi}^ndV_{\mc{S}_s}=\frac{2}{s}dV_{\mc{S}_s}, 
\\& \frac{d}{ds}tr\chi^L=-tr\chi tr\ul{\chi}+2\mc{K}-2|\zeta|^2-2div\zeta-\kappa tr\chi.
\end{split} 
 \eeq

Now, let us study the evolution  of the Hawking mass of such a foliation. Recall the 
mass aspect function (\ref{maspect}) and its relation (\ref{Hawk-MA})  with the Hawking mass of each $\mc{S}_s$. 
Now, using these and the evolution equations, along with the fact that 
$tr\ul{\chi}[\mc{S}_s]=\frac{2}{s}$, 
 we derive:

\beq
\label{monoton.deriv}
\begin{split}
&\frac{d}{ds}m_{\rm Hawk}[\mc{S}_s]=\frac{d}{ds}\big{(}\frac{r[\mc{S}_s]}{8\pi}
\int_{\mc{S}_s}\mu dV_s\big{)}
\\&=\frac{1}{2}\ol{tr\ul{\chi}}m_{\rm Hawk}[\mc{S}_s]-\frac{r[\mc{S}_s]}{16\pi}
\int_{\mc{S}_s}tr\ul{\chi}\mu dV_s+\frac{r[\mc{S}_s]}{32\pi}\int_{\mc{S}_s}tr\chi|
\hat{\ul{\chi}}|^2+ tr\ul{\chi}|\zeta|^2dV_s. 
\end{split}
\eeq
Now, first writing $\mu=\ol{\mu}+(\mu-\ol{\mu})$ and then recalling that 
$tr\ul{\chi}=\ol{tr\ul{\chi}}$,
 we finally obtain (\ref{monoton}). $\Box$
 
 As we will see in the next two subsections, all the hypersurfaces $\lmcN_{v,\tau}$ that
we  consider here 
 admit a luminosity foliation, and moreover the Hawking mass 
 is monotone increasing along such a foliation. It is in proving this latter property 
 that the assumption of closeness to the Schwarzschild solution is employed in an essential way.

 
\subsection{Construction of constant luminosity foliations, and their asymptotic behaviour.}

We now show how to construct a luminosity foliation on  the 
 surfaces $\lmcN_{v,\tau}$,  using their affine parameters $\la_{v,\tau}$ as a point of
  reference.
The construction here essentially follows 
\cite{Sauter}, whose notation we also adopt. 

Recall that the affine vector field $\ul{L}_{v,\tau}$ 
is normalized so that: $tr\ul{\chi}^{\ul{L}_{v,\tau}}[\mc{S}^\p_{v,\tau}]=2$ on 
$\mc{S}^\p_{v,\tau}$.
 $\lambda_{v,\tau}$ is the corresponding affine parameter, 
 with $\lambda|_{\mc{S}'_{v,\tau}}=1$. We refer to $\la_{v,\tau}$ as the ``background
 affine parameter'' on $\lmcN_{v,\tau}$. 
 
 To stress that this construction can be performed separately on each $\lmcN_{v,\tau}$, we
  use subscripts $v,\tau$ on all relevant quantities below, except for the luminosity parameters.
 These we still write as  $s$ instead of $s_{v,\tau}$ for notational convenience. (Note that
  each $s_{v,\tau}$ lives over $\lmcN_{v,\tau}$). In the subsequent sections where we study 
  the variation of $s_{v,\tau}$ under changes in $\tau$, we will use $s_{v,\tau}$.

The function $s$ that we are then seeking (on each 
null geodesic on $\lmcN_{v,\tau}$), can be encoded in a function $w_{v,\tau}(s,x)$ defined via the relation: 

\beq
\label{law}
\lambda_{v,\tau}=w_{v,\tau}(s,x).
\eeq
 
 The requirement (\ref{lumin}) can then be re-expressed as: 
 
 \beq
 \label{reqrmnt}
\frac{\partial}{ \partial_s}w_{v,\tau}(s,x)=\frac{2}{s\cdot tr\ul{\chi}^{\lambda_{v,\tau}}
(w_{v.\tau}(s,x),x)} 
 \eeq
 With the initial condition $w_{v,\tau}(1, x)=1$. 
 \newline
 
  In order to study the local and global existence  of a  solution to the above, it is useful to recall Lemma 
  \ref{trulchi} on the the asymptotic behaviour 
  of $tr\ul{\chi}^{\lambda_{v,\tau}}(\lambda,x)$ and its derivatives with respect to $x$.
In particular we recall that  
 \beq
 \label{asympt}
 tr\ul{\chi}^{\lambda_{v,\tau}}(x) =\frac{2}{\lambda_{v,\tau}}
 +\frac{a_{v,\tau}(x)}{\lambda_{v,\tau}^2},
 \eeq
where $a_{v,\tau}\in O^\delta_2(1)$,  and thus for some $0<c<C$
 \beq
 \label{bds}
  c\le \lambda tr\ul{\chi}^{\lambda_{v,\tau}} (x)\le C,
  \eeq
  for all $(x,\la)\in \lmcN$.
 Equation \eqref{reqrmnt} has a local solution in $s$. The equation 
   (\ref{reqrmnt}) implies that $w_{v,\tau}(s,x)$
 is increasing in $s$; the bounds (\ref{bds}) imply that the solution exists for all $s\ge 1$
  (for each $v, \tau$). 

We next claim: 

\begin{lemma} 
\label{lumin1}
Let $\lmcN_{v,\tau}$ be as in Definition \ref{Nvt}
 and let $\la_{v,\tau}$ 
be the background affine parameter, and $s$ the luminosity parameter constructed above.
We claim that
\beq
\label{addition}
(\la_{v,\tau}-1)^{-1}[s\la_{v,\tau}^{-1}-1]\in O^\delta_2(\la_{v,\tau}^{-1}),
\eeq
and moreover that   there exists a $\mc{C}^2$ function $\varphi_{v,\tau}(\phi,\theta)$ so
 that: 

\beq
\label{s-to-la}
s\la_{v,\tau}^{-1}=e^{\varphi_{v,\tau}(\phi,\theta)}+O^\delta_2(\la_{v,\tau}^{-1}).
\eeq
\end{lemma}

Postponing the proof of the above for a moment, we define:
\begin{definition}
\label{def-relation}
Consider 
the new (affine) vector field $\tilde{\ul{L}}_{v,\tau}:=e^{\varphi_{v,\tau}}\ul{L}_{v,\tau}$
and let $\tilde{\lambda}_{v,\tau}$ be its corresponding affine parameter.
\end{definition}
Observe that \eqref{s-to-la} implies:

\beq\label{relation}
s_{v,\tau} \tilde{\lambda}^{-1}_{v,\tau}=1+O^\delta_2(\la^{-1}_{v,\tau}). 
\eeq
 In particular, the spheres $\mc{S}(\{s_{v,\tau}=B \}\subset \lmcN_{v,\tau})$
  agree asymptotically (to leading order) with 
 the spheres $\mc{S}(\{\tilde{\lambda}_{v,\tau}=B\})\subset\lmcN_{v,\tau}$.
\newline

It will be necessary to calculate the \emph{first variation} $\dot{\varphi}_v$
of the functions $\varphi_{v,\tau}$ in $\tau$, which is defined via:

\beq
\label{varphi-dot}
\dot{\varphi}_v:=\frac{d}{d\tau}|_{\tau=0}\varphi_{v,\tau}.
\eeq
We claim:
\begin{lemma}
\label{lumin.var}
Consider  a smooth 1-parameter
family of null surfaces $\lmcN_{v,\tau}$ (emanating from spheres $\mc{S}_{v,\tau}$)
to which the assumption of Lemma \ref{trulchi.var}
applies. Then 
there exist functions $f^1(\la)\in O^1_2(1),
 f^2(\la)\in O^\delta_2(\la^{-2})$
 so that
\beq
\dot{\varphi}_v(\phi,\theta)=lim_{s\to\infty} f^1(s)\int_1^s f^2(t) \dot{a}_v(t)dt.
\eeq 
\end{lemma}

{\it Proof of Lemmas \ref{lumin1}, \ref{lumin.var}:} We prove both Lemmas together. 
Recall the parameters 
 $w_{v,\tau}(\la)$, $a_{v,\tau}(\la)$ on  $\lmcN_{v,\tau}$ as in \eqref{law}, 
 \eqref{asympt}.
 
 We now derive the asymptotic behaviour of the solution $w_{v,\tau}$ of (\ref{reqrmnt}), 
 for given $v,\tau$: 
 Given (\ref{asympt})  our equation becomes: 

\beq
\label{weq}
\partial_sw_{v,\tau}=\frac{w_{v,\tau}}{s(1+\frac{a_{v,\tau}(w_{v,\tau})}{2w_{v,\tau}})}.
\eeq
Letting $\tw_{v,\tau}:=\frac{w_{v,\tau}}{s}$ we transform the above into
 a new equation on $\tilde{w}_{v,\tau}$:

\beq
\label{new}
\partial_s\tw_{v,\tau}=-\frac{a_{v,\tau}(\tw_{v,\tau}s)\tw_{v,\tau}}{s[2s\tw_{v,\tau}+a_{v,\tau}(\tw_{v,\tau}s)]}.
\eeq

In view of the smallness of $a_{v,\tau}$, and since $\tla_{v,\tau}=1$ at $s=1$, 
a simple bootstrap argument reveals that $\tw_{v,\tau}$ stays $\delta$-close to 1 
for all $s\ge 1$. 
Then just integrating the above equation shows that $\tw_{v,\tau}$ converges to a limit 
as $s\to\infty$.
\beq
\tw_{v,\tau}(s)\to \tw_{v,\tau}(\infty),
\eeq 
with $|\tw_{v,\tau}(\infty)-1|=O(\delta)$ and $|\tw_{v,\tau}(s)-\tw_{v,\tau}(\infty)|\le Cs^{-1} $.
 In particular, we can define a continuous  function $\varphi_{v,\tau}(s,\phi,\theta)$ via:
\beq 
\label{wphi}
 \tw_{v,\tau}(s,\phi,\theta)=e^{\varphi_{v,\tau}(s,\phi,\theta)}.
 \eeq
 The above combined show \eqref{addition} and \eqref{s-to-la},\footnote{For the former 
 merely recall that 
 $\la=s$ on this initial sphere $\mc{S}_{v,\tau}$.} except for the angular regularity.
We let $\varphi_{v,\tau}(\phi,\theta):=lim_{s\to\infty} \varphi_{v,\tau}(\phi,\theta)$.
To obtain the extra regularity in the angular directions $\phi^1, \phi^2$, 
we just take $\partial_\phi,\partial_\theta$ derivatives of
\eqref{new}, up to two times. Then the claim on the extra regularity follows immediately from 
the resulting (linear) ODE, 
using the bounds we have derived in Lemma \ref{trulchi} on the angular derivatives of 
$a_{v,\tau}$. This proves Lemma \ref{lumin1}.
\newline

The proof of Lemma \ref{lumin.var} follows by just differentiating in $\tau$ 
equation (\ref{weq})
to derive (letting $\dot{a}_v(\la)$ stand for $\partial_\tau|_{\tau=0}a_{v,\tau}(\la)$ 
and $a'_{v,\tau}(\cdot)$ stands
 for the regular derivative in $\la$ of $a_{v,\tau}(\la)$). 
 
\beq
\partial_s\dot{\tw}_v=-\frac{\dot{a}_v(\tw_0 s)+a'_{v,0}(\tw_0 s)
\dot{\tw}_v s}{2s^2(1+\frac{a_{0}(w)}{2\tw_{0} s})}+
\frac{a_{0}(\tw s)}{2s^2(1+\frac{a_{0}(w)}{2\tw_{0} s})^2}(\dot{a}_v(\tw_{0} s)
+a'_{0}(\tw_{0} s)\dot{\tw}_{v} s).
\eeq
(Here $w=\tilde{w} s$ 
 lives over  the original surface $\lmcN$).  
Thus, seeing the above as a first-order linear ODE in $\dot{\tw}_v(s)$ we derive: 

\beq
\dot{\tw}_v(s)=O^\delta_2(1)\int_1^s O^\delta_2(t^{-2})\dot{a}_v(w_0(t))dt.
\eeq
By the definition \eqref{wphi} of $\varphi_{v,\tau}(s,\phi,\theta)$
 and passing to the limit $s\to\infty$, our 
claim follows.
 $\Box$
\newline

We have thus derived that any luminosity foliation on any $\lmcN_{v,\tau}$ is 
asymptotically equivalent to an \emph{affine} foliation of the 
same null hypersurface $\lmcN_{v,\tau}$. The relation between the luminosity parameter $s$ 
and the 
new affine parameter is given by \eqref{relation}.
\newline

As we have seen in the introduction, the main issue in capturing the Bondi energy at a section 
of $\mc{I}^-$ is 
to approximate that section by spheres that become asymptotically round. With that in mind, 
we  introduce a definition:

\begin{definition}
\label{renorm.GC}
Given any $\lmcN_{v,\tau}$ and either an affine parameter $\la$ or 
the luminosity parameter $s$, we let:  

\beq
\begin{split}
\label{GC}
&\mc{K}^{\infty,\la}_{v,\tau}(\phi,\theta):=  lim_{B\to\infty} B^2\mc{K}[\mc{S}(\{\la=B\})]
(\phi,\theta), 
\\&\mc{K}^{\infty,s}_{v,\tau}(\phi,\theta):=lim_{B\to\infty}
 B^2 \mc{K}[\mc{S}(\{s=B\})](\phi,\theta).
 \end{split}
\eeq
\end{definition}
Regarding the first limit, recall the discussion after \eqref{metric.infty} 
on the existence of a limit of 
the (renormalized) Gauss curvatures for an \emph{affine} foliation. Regarding the second limit, 
note that  we have not yet derived 
 its existence  at this point.

  We next claim that the (renormalized) 
 limits of the Gauss curvatures and the Hawking masses of the two foliations
 by $\tla_{v,\tau}$ and $s$ as in Definition \ref{def-relation}.
 agree. 
 This will enable us to replace luminosity foliations with suitable {\it  affine} foliations on 
 each of the null surfaces $\lmcN_{v,\tau}$ that we are considering.

 \begin{lemma}
 \label{G.curvatures}
 In the notation above we claim that on each $\lmcN_{v,\tau}$:
 
 \beq
 \label{arik1}
 \mc{K}_{v,\tau}^{\infty,s}(\phi,\theta)= \mc{K}_{v,\tau}^{\infty,\tla_{v,\tau}}(\phi,\theta)
 \eeq
  and 
 \beq
 \label{arik2}
 lim_{B\to\infty}m_{\rm Hawk}[\mc{S}(\{s=B\})]=
lim_{B\to\infty} m_{\rm Hawk}[\mc{S}(\{\tla_{v,\tau}=B\})]=0.
 \eeq
 \end{lemma}
{\it Proof:} To derive the first formula, we note that by the 
definition of $\ul{\chi}^{\ul{L}}_{ab}(\la)$, for \emph{any} affine vector field $\ul{L}$ 
with corresponding affine parameter $\la$:

\beq
\label{two.gs}
\mind_{ab}(\{\la=r_1(\phi,\theta)\})-\mind_{ab}(\{\la=r_2(\phi,\theta)\})=
2\int_{\{\la=r_1(\phi,\theta)\}}^{\{\la=r_2(\phi,\theta)\}}\ul{\chi}^{\ul{L}}_{ab}(\la),
\eeq
where 
the indices $a,b$ are assigned values from among the vector fields 
$\partial_\phi,\partial_\theta$.

Then, we choose the affine parameter $\tla_{v,\tau}$, and choose
$r_1(\phi,\theta)=B$. We also choose $r_2(\phi,\theta)$ to be the function so 
that:
\[
\{\tla_{v,\tau}=r_2(\phi,\theta)\}=\{s_{v,\tau}=B\}.
\]
Then  invoking the asymptotics \eqref{asymptotics1} of $\ul{\chi}$,  equations \eqref{two.gs} and
 \eqref{relation}, along with expression of the Gauss curvature in terms of second 
 coordinate derivatives of the metric of the spheres, we derive \eqref{arik1}.

To derive \eqref{arik2}
 we recall a formula for the Hawking mass:\footnote{ 
The $\chi, \ul{\chi}$ 
(along with their traces and trace-less parts) appearing below are defined 
relative to {\it any} pair of null vectors $L,\ul{L}$ 
 which are normal
to $\mc{S}$ and normalized so that $g(L,\ul{L})=2$.} 
 \beq
 \label{Hawk}
m_{\rm Hawk}[\mc{S}]=r[\mc{S}]\int_{\mc{S}} -\rho -\frac{1}{2}\hat{\chi}\cdot \hat{\ul{\chi}}-div\zeta dV_{\mc{S}}.
 \eeq
To derive this, we have used \eqref{maspect}, \eqref{Hawk-MA} and: \beq
 \label{Gauss}
 \mc{K}[\mc{S}]=-\rho-\frac{1}{2}\hat{\chi}[\mc{S}]\hat{\ul{\chi}}[\mc{S}]+\frac{1}{4}tr\chi[\mc{S}] tr\ul{\chi} [\mc{S}]
 \eeq 

Thus, to prove \eqref{arik2},  the main challenge is for any fixed large $B>0$ to compare 
 $tr\chi, \hat{\chi}$ on the spheres $\{s=B\}$ 
 and $\{\tilde{\lambda}=B\}$. 
 We will be using the formulas derived in section 2 of \cite{AlexShao2}, 
for the distortion function:
\beq
\label{new1}
e^{\psi_B}=\tilde{\lambda}(\{s=B\})B^{-1}.
\eeq
 ($\psi_B$ is defined to be constant on the null generators of $\lmcN_{v,\tau}$).
By \eqref{relation} we derive that: 

\beq
\label{new2}
\psi_B=O^\delta_2(B^{-1})
\eeq

Using \eqref{two.gs} 
we now compare the metric elements of the two spheres $\{s=B\}$ 
 and $\{\tilde{\lambda}=B\}$, via the natural map 
 that identifies points on the same null generator of $\lmcN$:
 \beq
 \gamma_{\{\tilde{\lambda}=B\}}=(1+O(B^{-1}))\gamma_{\{s=B\}}.
 \eeq 
A consequence of this is a comparison of the area elements and the areas of 
$\{s=B\}$ 
 and $\{\tilde{\lambda}=B\}$:
 \beq\label{areas-near}
 {\rm Area}[\{s=B\}]=(1+O(B^{-1})){\rm Area}[\{\tilde{\lambda}=B\}].
 \eeq

On the other hand, the transformation  laws of section 2 in \cite{AlexShao2} imply that:

\beq
\label{rhos-near}
B^2\rho(\{s=B\}),B^2\rho(\{\tilde{\lambda}=B\})\}=O(B^{-1}).
\eeq
Also, the transformation laws of subsection \ref{transf-AS3} in 
the present paper  imply
\beq
\label{trchis-near}
B^2\{tr\chi[\mc{S}(\{s=B\})]tr\ul{\chi}[\mc{S}(\{s=B\})]- tr\chi[\mc{S}(\{\tilde{\lambda}=B\})]tr\ul{\chi}[\mc{S}(\{\tilde{\lambda}=B\})]\}=O(B^{-1}).
\eeq
\beq
\label{hatchis-near}
B^3\{\hat{\chi}[\mc{S}(\{s=B\})]\hat{\ul{\chi}}[\mc{S}(\{s=B\})] -\hat{\chi}[\mc{S}(\{\tilde{\lambda}=B\})]\hat{\ul{\chi}}
[\mc{S}(\{\tilde{\lambda}=B\})]\}=O(B^{-1}).
\eeq
 These equations, combined with \eqref{Hawk}  prove our claim. $\Box$
 \newline
 
 \begin{remark}
 \label{zeta-chi-also}
 Note that the transformation laws \eqref{ulchi-trans}, \eqref{zeta-trans}, \eqref{chi-trans}
 invoked above show that since  the Ricci 
 coefficients 
 $\zeta$,
 $\chi$ associated with the affine parameter $\tla_{v,\tau}$ satisfy the bounds \eqref{zet.chi} of Lemma 
 \ref{zeta.chi}, then the same bounds are satisfied by the Ricci coefficient associated with 
 the luminosity parameter $s$.  
 \end{remark}

 We next prove that the luminosity foliations on all the hypersurfaces $\lmcN_{v,\tau}$
 have the desired monotonicity of the Hawking mass: 
 \newline
 
 {\bf Monotonicity of the Hawking mass for luminosity foliations:}

 \begin{lemma}
 \label{monotonicity-forall}
 Consider any hypersurface $\lmcN_{v,\tau}$ as in Definition \ref{Nvt}, and 
 let $\{\mc{S}_s\}_{s\ge 1}$ be its luminosity foliation. Then $tr\chi^{L_s}[\mc{S}_s]>0$ for 
 all 
 $s>1$. (Here $L_s$ is the future-directed outgoing null normal to $\mc{S}_s$). 
 In particular, in view of \eqref{monoton}, $m_{\rm Hawk}[\mc{S}_s]$ is an 
 increasing function in $s$. 
 \end{lemma}
 {\it Proof:} We show this in two steps. Firstly, observe that the level sets 
 of the original affine parameter $\la_{v,\tau}$ satisfy $\tr\chi^{L_{v,\tau}}
 [\mc{S}_{\la_{v,\tau}}]>0$. This follows from \eqref{lower.bd} which yields a positive lower bound for 
$\tr\chi^{L_{v,\tau}}
 [\mc{S}_{\la_{v,\tau}}]$. 
 Secondly we use this lower bound to derive $tr\chi^{L_s}[\mc{S}_s]>0$.
 This second claim in fact follows straightforwardly from \eqref{lower.bd}, 
 coupled with \eqref{addition} (which encodes how the spheres $\mc{S}_{\{s=B\}}$ 
 are small
 perturbations of the spheres $\mc{S}_{\{\la_{v,\tau}=B\}}$),
 and the transformation law \eqref{eq.cf_chibar}. $\Box$

\section{Variations of the null surfaces  and their luminosity foliations.}
\label{sec-variations}

 We next seek to capture how a variation $\lmcN_{v,\tau}$ of the original $\lmcN_0$ induces a variation on the Gauss curvature
of the metric at infinity of $\lmcN_{v,\tau}$ associated with the luminosity 
foliations on these surfaces. (See the first equation in \eqref{GC}). 

\subsection{Varying null surfaces and  luminosity foliations: The effect on the Gauss curvature at infinity.}

We consider  the family $\lmcN_{v,\tau}$ of smooth null surfaces 
in Definition \ref{Nvt}. 
We also consider the associated functions
$w_{v,\tau}(\phi,\theta,s)$ and let $s_{v,\tau}$ to be the luminosity parameters on $\lmcN_{v,\tau}$. 
(We write out $s_{v,\tau}$ instead of just $s$, to stress that
 we are studying the variation of $\lmcN_{v,\tau}$ and
the parameters  $s_{v,\tau}$ defined over them).

Our goal in this section is to calculate the first variation  of the renormalized 
 Gauss curvatures $\mc{K}^{\infty,s}_{v,\tau}$, around $\mc{K}_{0}^{\infty,s}$ (see 
 \eqref{GC}) which corresponds to the initial null hypersurface $\lmcN_0$.
 In other words we seek to 
capture:
\beq
\label{1st.var.goal}
\frac{d}{d\tau}|_{\tau=0} lim_{B\to\infty} B^2\mc{K}[\{s_{v,\tau}=B \}\subset\lmcN_{v,\tau}].
\eeq

\begin{remark}
\label{generalization}
We note for future reference that the precise same calculation can be applied also to 
capture the 
first variation of Gauss curvatures around \emph{any} null surface $\lmcN_\omega$ with 
$||\omega||_{W^{4,p}(\mc{S}_0)}\le 10^{-1}m$. 
This  follows readily in view of the assumed bounds on the curvature on the surfaces $\lmcN_\omega$. 
In particular all the formulas  we derive remain true, by just replacing 
the Ricci coefficients and curvature components on $\lmcN_0$ by those on $\lmcN_\omega$.  
\end{remark}

 \begin{definition}
 On each $\lmcN_{v,\tau}$ we let $s_{v,\tau}$ be the luminosity parameter. 
 We let $\mc{S}_{v,\tau}[B]$ be the level set 
 $\{s_{v,\tau}=B\}\subset\lmcN_{v,\tau}$ and $\gamma_{v,\tau}[B]$ the induced metric on
 this  sphere. We then let:
 \beq
 \gamma^\infty_{v,\tau}:=lim_{B\to\infty} B^{-2}\gamma_{v,\tau}[B]
 \eeq
 The limit is understood in the sense of components relative to the coordinate vector fields 
 $\partial_\phi,\partial_\theta$.
For $\tau=0$ we just denote the corresponding limit metric by $\gamma^\infty$.
 \end{definition}
 (Note that these limits exist, by combining \eqref{relation} with \eqref{metric.infty}, 
 to derive that the corresponding limit exists for the level sets of the \emph{affine} 
 parameter $\tla_{v,\tau}$). Note further (as mentioned above) that the same equations imply: 
 
 \beq
\label{GC,lims} 
\mc{K}[  \gamma^\infty_{v,\tau}]=lim_{B\to\infty} B^2\mc{K}[\{s_{v,\tau}=B \}].
 \eeq

To capture the variation of the Gauss curvatures, we proceed in two steps: 
The variation of the null surfaces $\lmcN_{v,\tau}$, foliated by the backgound
 \emph{affine} parameters $\la_{v,\tau}$  is encoded 
in the Jacobi fields (\ref{Jac}) along the 2-parameter family of geodesics 
$\gamma_Q\subset \lmcN_{v,\tau}$. 
Recall the \emph{new} affine parameters $\tla_{v,\tau}$ defined in Definition \ref{def-relation}, 
which asymptote to the luminosity foliations of the hypersurfaces $\lmcN_{v,\tau}$.
We then define the family of \emph{modified} Jacobi fields $\tilde{J}$ 
 that correspond to these affine parameters: 

\beq
\label{mod.Jac}
\tilde{J}_{v,Q}(B):= \frac{d}{d\tau}|_{\tau=0} \{\tla^Q_{v,\tau}=B\}
\eeq

We consider  $\tJ_v$ expressed
in the frame $\ul{L}, e^1, e^2, L$. as in Definition \ref{frame-constr}:

\begin{definition}
\label{1st.defn}
 We denote the components of the Jacobi fields $\tJ_v$ expressed
 with respect to the above frame by $J_v^{\ul{L}}, J_v^A, A=1,2$ and $J_v^L$. 
 We will think of these components with respect to the background affine parameter $\la$ on $\lmcN$. The \emph{prime} $\prime$ will stand for the derivative with respect to $\la$. 
 (In particular $(\tJ^{\ul{L}}_v)^\prime:=\frac{d}{d\la}\tJ^{\ul{L}}_v(\la)$. 
\end{definition}

We let:
\[
(\tJ_v^{\ul{L}})^\p_\infty:=lim_{\la\to\infty} (\tJ_v^{\ul{L}})^\p(\la).
\] 
(The existence of this limit will be derived below, for every $v\in W^{4,p}(\mc{S}_0)$).

We claim that:

\begin{proposition}
\label{propo.var}
With the identification of coordinates described above:

\beq
\label{main-varn-claim}
\mc{K}[\gamma^\infty_{v,\tau}]-\mc{K}[\gamma^\infty]=2\tau [\Delta_{\gamma^\infty}+2\mc{K}(\gamma^\infty)]
((\tJ_v^{\ul{L}})^\prime_\infty)+o(\tau).
\eeq
\end{proposition}

In fact, using the definition \ref{def-relation} of $\tla_{v,\tau}$ along with \eqref{varphi-dot}, 
we find readily that:
\beq
\label{important}
(\tJ_v^{\ul{L}})^\prime(\la)=(J_v^{\ul{L}})^\prime(\la)+\dot{\varphi}_v(\la).
\eeq
The evaluation of the two terms in the RHS of the above will 
be performed in the next section. For now, we prove the Proposition above: 
\newline

{\it Proof of Proposition \eqref{propo.var}:} The key insight behind the proof is that 
the variation of the spheres under study can be decomposed into one \emph{tangential} 
to the original null surface $\lmcN_0$ and one \emph{transverse} to it. We find that the transverse component 
of the variation only contributes an error term to the variation of the (renormalized) Gauss curvature. 
On the other hand, the tangential variation induces a (linearized) \emph{conformal}
 change of the underlying metric, since it corresponds 
(up to error terms) to a first variation of affine foliations. In a different guise, this latter fact was also used in \cite{AlexShao3}
(albeit on a single, un-perturbed null hypersurface); as noted there, 
the intuition behind this goes back to the ambient metric construction of Fefferman and Graham \cite{FG}. 

 Returning to the proof, observe that it suffices to show that:

\beq
\label{main-varn-claim2}
B^2\mc{K}[\{s_{v,\tau}=B \}]-B^2\mc{K}[\{s_0=B \}]=2\tau 
B^2[\Delta_{\gamma[B]}+2\mc{K}(\gamma[B])]
[(\tJ^{\ul{L}}_v)^\prime]+o(\tau)+O(B^{-1}).
\eeq
 Invoking the limit 
$B^{-2}\gamma[B]\to \gamma^\infty$ (the convergence being in $\mc{C}^2$,
as noted in the proof of Lemma \ref{G.curvatures})
we note that:
\[
B^2[\Delta_{\gamma[B]}-\mc{K}(\gamma[B])]
(\tJ_v^{\ul{L}})^\prime\to [\Delta_{\gamma^\infty}-\mc{K}(\gamma^\infty)]
(\tJ_v^{\ul{L}})^\prime,
\]
with the convergence being in $L^p(\mc{S})$. Thus it suffices to show 
\eqref{main-varn-claim2} to derive \eqref{main-varn-claim}.

To capture the difference in the LHS of \eqref{main-varn-claim2}, we will proceed in
 six steps, suitably approximating the spheres

\[
\{s_{v,\tau}=B  \}\subset \lmcN_{v,\tau}, \{s_0=B\}\subset\lmcN_0.
\]

Recall that $\ul{L}_{v,\tau}$ is the affine vector field on $\lmcN_{v,\tau}$ 
normalized so that $tr\ul{\chi}^{\ul{L}_{v,\tau}}[\mc{S}^\prime_{v,\tau}]=2$. Recall 
that $\la_{v,\tau}$ is the corresponding affine parameter.
We also recall that $\tla_{v,\tau}$ is the affine parameter on $\lmcN_{v,\tau}$ which
 asymptotes to the luminosity parameter $s_{v,\tau}$; see \eqref{relation}.

\begin{definition}
We  let $\ul{L}^\flat_{v,\tau}$ be a new affine vector field on $\lmcN_0$ defined via:
\[
\ul{L}^\flat_{v,\tau}:= (1+\tau (\tJ^{\ul{L}})^\prime_\infty)\ul{L},
\]
and we let 
${\la}^\flat_{v,\tau}$ be the corresponding affine function over $\lmcN_0$,\footnote{Recall that 
 $\lmcN_0$ is the original null hypersurface. Thus these are 1-parameter families
 of affine  vector fields over the \emph{original} hypersurface.}
 normalized such that $\la^\flat_{v,\tau}=1$ on $\mc{S}_0$ and 
$\ul{L}^\flat_{v,\tau}({\la}^\flat_{v,\tau})=1$.

Let $L'_{v,\tau}$ be a null vector field, 
 normal to the level sets of ${\la}^\flat_{v,\tau}$ on $\lmcN_0$, 
 with\\ $g(L'_{v,\tau},\ul{L}^\flat_{v,\tau})=2$.
\end{definition}


We then let:
\begin{definition}
\label{six.approxns}
\begin{enumerate}
\item $\mc{S}^1_{0}(B):=\{ s_{0}=B\}\subset\lmcN_0$.
\item $\mc{S}^2_{0}(B):=\{ \tla_{0}=B\}\subset\lmcN_0$.
\item $\mc{S}^3_{v,\tau}(B):=\{ \la^\flat_{v,\tau}=B\}\subset\lmcN_0$.
 \item $\mc{S}^4_{v,\tau}(B)$ is  the sphere obtained from $\mc{S}^3_{v,\tau}(B)$ 
 by flowing along the geodesics emanating from the vector field $L^\prime_{v,\tau}$ by 
 $J^{L\rq{}_{v,\tau}}\tau$ 
in the corresponding affine parameter.
 \item $\mc{S}^5_{v,\tau}(B):=\{ \tla_{v,\tau}=B\}\subset\lmcN_{v,\tau}$.
 \item $\mc{S}^6_{v,\tau}(B):=\{s_{v,\tau}=B\}\subset \lmcN_{v,\tau}$.
\end{enumerate}
\end{definition}

Our aim is to show that 
\beq
\label{var.end}
B^2\mc{K}[\mc{S}^1_{0}(B)]-B^2\mc{K}[\mc{S}^6_{v,\tau}(B)],
\eeq
equals the RHS of \eqref{main-varn-claim}
(up to errors of the form $O(B^{-1})$, plus $o(\tau)$).
 (The difference in \eqref{var.end} is taken
in the coordinates $\phi,\theta$ defined over both spheres 
$\mc{S}^1_0(B), \mc{S}^6_{v,\tau}(B)$).
\newline

To do this, we seek to ``move'' from the first to the sixth sphere by successively moving along 
the intermediate spheres defined above.

Invoking Lemma \ref{G.curvatures}, we find that:

\beq
\label{easy1}
|B^2\mc{K}[\mc{S}^1_{0}(B)]-B^2\mc{K}[\mc{S}^2_{0}(B)]|=O(B^{-1}), 
|B^2\mc{K}[\mc{S}^5_{v,\tau}(B)]-
B^2\mc{K}[\mc{S}^6_{v,\tau}(B)]|=O(B^{-1}).
\eeq

Next, our Assumption \ref{assumption} on the space-time implies that: 
\beq
\label{reg1}
 |B^2\mc{K}[\mc{S}^5_{v,\tau}(B)]- B^2\mc{K}[\mc{S}^4_{v,\tau}(B)]|=o(\tau)+O(B^{-1}).
\eeq

The above combined then show that: 

\beq
\label{important-again}
\begin{split}
&B^2\mc{K}[\mc{S}^6_{v,\tau}(B)]-B^2\mc{K}[\mc{S}^1_{v,0}(B)]=
B^2\mc{K}[\mc{S}^3_{v,\tau}(B)]-B^2\mc{K}[\mc{S}^2_{v,0}(B)]
\\&+B^2\mc{K}[\mc{S}^4_{v,\tau}(B)]-B^2\mc{K}[\mc{S}^3_{v,\tau}(B)]
+O(B^{-1})+o(\tau).
\end{split}
\eeq

Thus it suffices to estimate the two 
differences of the RHS of the above, up to errors of the form $O(B^{-1})$ and $o(\tau)$. 
\newline

We recall that the functions $tr\chi^{L'}$, 
$tr\chi^{L'_{v,\tau}}$ can be thought of as a scalar-valued functions over the null surface $\lmcN_0$;
 we can then consider their restriction to any sphere $\mc{S}^3_{v,\tau}$. 
This yields  a scalar-valued functions over that sphere. 

 
With this convention, 
recalling  \eqref{new.trchi} we derive that:

\beq
\begin{split}
&B\cdot tr\chi^{L'_{v,\tau}}[\mc{S}^3_{v,\tau}(B)]=
B(1+2\tau(\tJ^{\ul{L}}_v)^\prime_\infty )tr\chi^{L'}[\mc{S}^2_{0}(B)]
+2\tau B^2\Delta_{\gamma_{\mc{S}^2_{0}(B)}}
(\tJ^{\ul{L}}_v)^\prime_\infty
\\&+o(\tau)+O(B^{-1}).
\end{split}
\eeq
Now, we recall that $\gamma^\infty$ is 
the metric induced ``at infinity'' on $\lmcN_0$ by the affine parameter 
$\tla_{0}$. We apply formulas \eqref{Gauss}, \eqref{trulchi} and recall the decay 
of the component $\rho$ in \eqref{rhos-decay}
 to the affine parameter $\tla_0$ to derive that:
\beq
 Btr\chi^{L'}[\mc{S}^2_{0}(B)]
 =2\mc{K}[\gamma^\infty]+O(B^{-1}).
\eeq 
Combining the above two equations,   we derive that: 
\beq
 B^2\mc{K}[\mc{S}^3_{v,\tau}(B)]=B^2\mc{K}[\mc{S}^2_{0}(B)]+
2\tau [\Delta_{\gamma^\infty} +2\mc{K}[\gamma^\infty]](\tJ^{\ul{L}}_v)^\p
+O(B^{-1})+o(\tau).
\eeq
 
Now we proceed to show
\beq
\label{iapwn}
B^2\mc{K}[\mc{S}^4_{v,\tau}(B)]-B^2\mc{K}[\mc{S}^3_{v,\tau}(B)]=O(B^{-1})+o(\tau):
\eeq

  Let $\mc{H}_{v,\tau}$ be the (small, incomplete) 
future-directed null surface emanating from $\mc{S}^3_{v,\tau}$ in the direction of 
$L'_{v,\tau}$. 
Let $L'_{v,\tau}$ be the corresponding affine vector field on this surface, and 
$\ul{u}_{v,\tau}$ 
the corresponding affine parameter, with $\ul{u}_{v,\tau}=0$ on 
$\mc{S}^3_{v,\tau}$.
Let us denote by $\ul{L}^\sharp_{v,\tau}$ the transverse 
null vector field along $\mc{H}_{v,\tau}$
which is normal to the level sets of $\ul{u}_{v,\tau}$ so that 
$g(L'_{v,\tau}, \ul{L}^\sharp_{v,\tau})=2$.
We then observe that by definition:
\beq
\mc{S}^4_{v,\tau}:=\{\ul{u}_{v,\tau}=\tau \}\subset \mc{H}_{v,\tau}.
\eeq
This, implies that 

\beq
\label{evolutions-transverse}
\begin{split}
&B^2 \chi^{L'_{v,\tau}}_{ab}[\mc{S}^4_{v,\tau}(B)]
 (\ul{\chi}^{\ul{L}^\sharp_{v,\tau}})^{ab}[\mc{S}^4_{v,\tau}(B)]
-B^2\chi_{ab}^{L'_{v,\tau}}[\mc{S}^3_{v,\tau}(B)]
(\ul{\chi}^{\ul{L}^\sharp_{v,\tau}})^{ab}[\mc{S}^3_{v,\tau}(B)]
\\&=B^2\nasla_{L'_{v,\tau}}\{\chi^{L'_{v,\tau}}_{ab}[\mc{S}^3_{v,\tau}(B)]
 (\ul{\chi}^{\ul{L}^\sharp_{v,\tau}})^{ab}[\mc{S}^3_{v,\tau}(B)]\}+o(\tau)
 \\&=O(B^{-1})+o(\tau). 
\end{split}
\eeq
The last equation follows from the first and last formulas in \eqref{eq.structure_ev} and the bounds in 
Lemma \ref{zeta.chi}.

Furthermore, by our decay assumptions \eqref{curv.decay}
 on the curvature components (which are assumed on all $\lmcN_{v,\tau}$)
\beq
\label{rhos-decay}
B^2\rho[\mc{S}^4_{v,\tau}(B)], B^2\rho[\mc{S}^3_{v,\tau}(B)]=O(B^{-1}).
\eeq
Consequently, invoking equation \eqref{Gauss}, we derive \eqref{iapwn}. 

This completes the proof of Proposition \ref{propo.var}. $\Box$
\newline

Thus the main point that remains is to express $(\tJ^{\ul{L}}_v)^\prime_\infty$
 in terms of the function $e^v$. 
We take this up in the next subsection.

\subsection{Variations of the  null surfaces  and Jacobi fields.}

The goal for the remainder of this section is to prove:  
 \begin{proposition}
 \label{tJ-est}
 There exist \emph{fixed}  functions (i.~e.~independent of $v$) 
 $f^i\in \mc{C}^2(\mc{S}_0), i=0,1,2,3$
 with $|f^i|_{\mc{C}^2(\mc{S}_0)}\le\delta$
   so that:
 \beq
 \begin{split}\label{tJ-est-eqn}
& (\tJ^{\ul{L}}_v)^\p_\infty=(1+f^0)\{-2\Delta_{\mc{S}_0}e^v+ f^1\partial_1 e^v+f^2\partial_2 e^v
+f^3e^v\}.
\end{split}
 \eeq
  \end{proposition}
This will be proven by combining Propositions \ref{jacobi} and \ref{phi-var} below, 
see formula \eqref{important}. 

We proceed to show:

\begin{proposition}
\label{jacobi}
We claim that there exist fixed functions (i.~e.~functions independent of $v$)  
$k^1\in O^\delta_2(\la^{-1}), k^2\in O^\delta_2(\la^{-1}), 
\tilde{f}^{\ul{L}}\in O^\delta_2(\la)$, 
$\tilde{f}^1, \tilde{f}^2\in O^\delta_2(\la)$,\footnote{Recall that the vector 
fields $e_1, e_2$ below correspond to 
$\la^{-1}\partial_{\phi^1}, \la^{-1}\partial_{\phi^2}$.}  so that:

\beq
\label{jacobi-eq}
\begin{split}
&J^L=e^v, J^1=k^1e^v, J^2=k_2e^v, 
\\&  (J^{\ul{L}})^\p=-2\Delta_{\mc{S}_0}e^v+ \sum_{C=1}^2\tilde{f}^C e_C[e^v]
+\tilde{f}^{\ul{L}}e^v.
\end{split}
\eeq
Moreover, the limit $lim_{\la\to\infty}\la^{-1}J^{\ul{L}}(\la)$ exists, and is a 
$\delta$-small (in $\mc{C}^2$) perturbation of the operator $\Delta_{\mc{S}_0}[e^v]$. 
\end{proposition}

{\it Proof:} 
By the construction above, (\ref{formula.F})  and the definition (\ref{Jac}) we 
readily find that for each point $Q\in\mc{S}_0$ (with $\lambda^Q(Q)=1$): 

\beq
\begin{split}
&J^L(1)=e^v, J^1(1), J^2(1)=0, 
\\&J^{\ul{L}}(1)= e^v|\hat{\chi}|^2[\mc{S}_0]\{
-2\rho[\mc{S}_0]-2div\zeta [\mc{S}_0]-2|\zeta[\mc{S}_0]|^2+\hat{\chi}
[\mc{S}_0]\hat{\ul{\chi}}[\mc{S}_0]\}^{-1}
\end{split}
\eeq

Moreover, the requirement that $J$ corresponds to 
a variation by {\it null} geodesics implies that $g(\ul{L},\nabla_{\ul{L}}{J})=0$; this in turn 
forces $({J}^L)^\prime(1)=0$. We also have by construction $({J}^A)^\prime(1)=0$, for $A=1,2$.
 Lastly, we have  derived in (\ref{right.scale}) that 
\beq
\begin{split}
\label{Lbar1}
&(J^{\ul{L}})^\prime(1)=  -2 \Delta_{\mc{S}_0}e^v+ O^\delta_2(1) \cdot\nabla e^v-(2\rho[\mc{S}_0]+O^\delta_2(1) )e^v. 
\end{split}
\eeq






{\it Jacobi equations.}  
We study the Jacobi equation relative to the frame $\ul{L}, e_1, e_2, L$.
  We recall the Jacobi equation:
\beq
\label{gen.Jacobi}
(\nabla_{\ul{L}\ul{L}}{J})^B=-{R_{\ul{L}A\ul{L}}}^BJ^A.
\eeq

In order to solve for the Jacobi field, we first need to express the LHS of the above in terms of 
$\nabla_{\ul{L}}$ derivatives of the various components, relative to the 
frame $\ul{L}, L, e^1, e^2$. Recalling the evolution equations \eqref{frame.evol},
 \eqref{2forms}, \eqref{asymptotics1}
we calculate:\footnote{Recall that $\prime$ stands for the regular $\frac{d}{d\la}$-derivative
 of scalar-valued quantities. }
\beq
\begin{split}
&\nabla_{\ul{L}\ul{L}}J=\nabla_{\ul{L}\ul{L}}(J^{\ul{L}}\ul{L}+J^LL+
\sum_{A=1}^2J^Ae_A)
\\&=({J}^{\ul{L}})^{\p\p}\ul{L}+({J}^L)^{\p\p}L+
\sum_{A=1}^2({J}^A)^{\p\p}e_A+2[(J^L)^\p\nabla_{\ul{L}}L
+\sum_{A=1}^2(J^A)^\p\nabla_{\ul{L}}e_A]
\\&+
J^L\nabla_{\ul{L}}(\nabla_{\ul{L}}L)+\sum_{A=1}^2J^A\nabla_{\ul{L}}(\nabla_{\ul{L}}e_A)
\\&=({J}^{\ul{L}})^{\p\p}\ul{L}+({J}^L)^{\p\p}L+
\sum_{A=1}^2({J}^A)^{\p\p}e_A+2({J}^L)^\p\zeta^Ae_A
\\&+2\sum_{A=1}^2({J}^A)^\p
(-\frac{1}{\la}e_A+(\ul{\chi}_A^\sharp)^Be_B+\zeta_A \ul{L})
\\&+J^L\nabla_{\ul{L}}[\zeta^Ae_A]+
\sum_{A=1}^2J^A\nabla_{\ul{L}}
(-\frac{1}{\la}e_A+(\ul{\chi}_A^\sharp)^Be_B+\zeta_A \ul{L})
\\&=({J}^{\ul{L}})^{\p\p}\ul{L}+({J}^L)^{\p\p}L+
\sum_{A=1}^2({J}^A)^{\p\p}e_A+2({J}^L)^\p\zeta^Ae_A
\\&+2\sum_{A=1}^2({J}^A)^\p
(-\frac{1}{\la}e_A+(\ul{\chi}_A^\sharp)^Be_B+\zeta_A \ul{L})
\\&+J^L[\nabla_{\ul{L}}\zeta^A e_A-\frac{1}{\la}\zeta^Ae_A+
\zeta^A(\ul{\chi}^\sharp)_A^B
e_B+\zeta^A\zeta_A\ul{L}]
+\sum_{A=1}^2J^A
(   (\nabla_{\ul{L}}({\chi}_A^\sharp)^Be_B+{\zeta}^\p_A \ul{L})
\\&+
\sum_{A=1}^2J^A[\frac{2}{\la^2}e_A-\frac{2}{\la}  (\ul{\chi}_A^\sharp)^B
e_B-\la^{-1}\zeta_A \ul{L}+
(\ul{\chi}_A^\sharp)^B(\ul{\chi}_B^\sharp)^Ce_C+(\ul{\chi}_A^\sharp)^B\zeta_B \ul{L}]
\\&=(J^{\ul{L}})^{\p\p}\ul{L}+(J^L)^{\p\p}L+
\sum_{A=1}^2(J^A)^{\p\p}e_A+(J^L)^\p\zeta^Ae_A
\\&+
2\sum_{A=1}^2(J^A)^\p
((\frac{2a\mind_A^B+h_A^B}{2\la^2})e_B+\zeta_A \ul{L})
\\&+J^L[({\zeta}^A)^\p e_A-\frac{1}{\la}\zeta^Ae_A+\zeta^A(\ul{\chi}^B)_A
e_B+\zeta^A\zeta_A\ul{L}]
\\&+
\sum_{A=1}^2J^A
(   (\frac{{(a\mind_{AB})^\p}}{2\la^2}+\frac{({h}_A^B)^\p}{\la^2} )   e_B+
({\zeta}_A)^\p \ul{L})
\\&+\sum_{A=1}^2J^A[-\frac{2}{\la} 
 (\frac{a}{2\la^2}e_A+\frac{h_A^B}{\la^2}
e_B)+\frac{a^2}{4\la^4}e_A+
\frac{h_A^Bh_B^C}{\la^4}e_C+(\ul{\chi}_A^\sharp)^B\zeta_B \ul{L}]-
\frac{\zeta_A}{\la}\ul{L}
\end{split}
\eeq

In short,  using the equations \eqref{zet.chi} and \eqref{asymptotics1}, we derive 
the equation:

\beq
\label{Final}
\begin{split}
&\nabla_{\ul{L}\ul{L}}J=({J}^{\ul{L}})^{\p\p}\ul{L}+({J}^L)^{\p\p}L+
\sum_{A=1}^2({J}^A)^{\p\p}e_A+({J}^L)^\p\sum_{A=1}^2O^\delta_2(\la^{-2})
e_A
\\&+
\sum_{A,B=1}^2O^\delta_2(\la^{-2})({J}^A)^\p e_B
+\sum_{A=1}^2O^\delta_2(\la^{-2}) ({J}^A)^\p\ul{L}
\\& +\sum_{A=1}^2J^LO^\delta_2(\la^{-3})e_A  +
\sum_{A,B=1}^2O^\delta_2(\la^{-3})J^Ae_B
+\sum_{A=1}^2J^A O^\delta_2(\la^{-3}) \ul{L}
\end{split}
\eeq
Now,  setting $B=L$ in (\ref{gen.Jacobi}) we derive: 
  \beq
(\nabla_{\ul{L}\ul{L}}{J})^L={R_{a\ul{L}\ul{L}}}^LJ^a={R_{a\ul{L}\ul{L}\ul{L}}}J^a=0.
  \eeq
  Refer to equation \eqref{Final}. Observe that the coefficient of $L$ is precisely 
  $({J}^L)^{\p\p}$.
Thus we derive: 
\[
({J}^L)^{\p\p}(\la)=0
\]
Therefore, using the initial conditions
$J^L(1)=e^v$, $({J}^L)^\p(1)=0$ at $\la=1$ we find: 
\beq
\label{L.easy}
J^L=e^v
\eeq   
We now consider the $B$-components of the Jacobi equation, with $B=1,2$. Recalling that $J^L=e^v$ we derive: 

\beq
\label{b.cpnt}
\begin{split}
&({J}^B)^{\p\p}+\sum_{A=1}^2[O^\delta_2(\la^{-2})({J}^A)^\p+O^\delta_2(\la^{-3})J^A]+O^\delta_2(\la^{-3})e^v
\\&=-{R_{\ul{L}S\ul{L}}^B}J^S=R_{\ul{L}L\ul{L}b}J^L+R_{\ul{L}A\ul{L}}^BJ^A=
R_{\ul{L}L\ul{L}}^Be^v+R_{\ul{L}A\ul{L}}^BJ^A
\\&=\ul{\beta}^Be^v+\ul{\alpha}_{A}^BJ^A.
\end{split}
\eeq
Moreover, since (\ref{b.cpnt}) is a linear ODE with trivial initial data, (since 
$J^B(1)=({J}^B)^\p(1)=0$ for $B=1,2$ as noted above),
we derive that the solution $J^1, J^2$ depend 
{\it linearly} on the parameter $e^v$. In particular there exist two functions
 $\vartheta^1(\la),\vartheta^2(\la)$ so that: 
\beq
\label{solnform}
J^B(\la)=\vartheta^B(\la)e^v.
\eeq

Then, using the explicit form of the equation above to express it as a system of first order ODEs, we readily find that $\vartheta_b(\la)\in O^\delta_2(\la^{-1})$.

Finally considering the component $B=\ul{L}$ in the Jacobi equation, along with 
\eqref{Final}, \eqref{solnform}, $J^L=e^v$, we find:
\beq
\label{Jac3}
\begin{split}
&({J}^{\ul{L}})^{\p\p}+O^\delta_2(\la^{-3})e^v
\\&=-\sum_{S=1}^2{R_{\ul{L}S\ul{L}}}^{\ul{L}}J^S
-R_{\ul{L}L\ul{L}L}J^L=-\sum_{S=1}^2{R_{\ul{L}S\ul{L}}}^{\ul{L}}J^S-4\rho e^v 
\end{split}
\eeq

Using    \eqref{curv.decay}, along with the fact that $u=2m\la$ by construction,  we derive:
\beq
({J}^{\ul{L}})^{\p\p}(\la)=
[\frac{8m}{u^3}+O^\delta_2(\la^{-3})]e^v=[\frac{1}{m^2\la^3}+O^\delta_2(\la^{-3})]e^v.
\eeq

Integrating the above in $\la$  we derive:

\beq
({J}^{\ul{L}})^{\p}(\la)=(J^{\ul{L}})^\p(1)+\frac{1}{m^2}\int_1^\la t^{-3}dt +O^\delta_2(\la^{-2}). 
\eeq
Therefore using the 
 initial conditions \eqref{Lbar1}, we find that: 
\beq
\label{limit.J}
lim_{\la\to\infty}(J^{\ul{L}})^\prime(\la)=-2\Delta_{\mc{S}_0}e^v +\sum_{A=1}^2
O_2^\delta(1)\partial_{\phi^A} e^v+
O_2^\delta(1)e^v. 
\eeq
This precisely yields  
the last equation in \eqref{jacobi-eq}. $\Box$


  \subsection{Jacobi fields and the first variation of Weyl curvature.}

Recall \eqref{varphi-dot}. 
Our aim here is to show that:

\begin{proposition}
\label{phi-var}
There exist functions $n^1(\phi,\theta), n^2(\phi,\theta), n^3(\phi,\theta)\in O^\delta_2
(\mc{S}_\infty)$
so that:\footnote{Recall that $O^\delta_2$ is a slight abuse of notation and refers 
to functions defined over $\mc{S}_0$ (or $\mc{S}_\infty$, via the identification of points on
 the same null generators on $\lmcN_0$). It requires the function and to rotational derivatives
  $\partial_{\phi^i}$ to be bounded by $\delta$.}
\[
\dot{\varphi}_v=n^1\cdot [\Delta_{\mc{S}_0}e^v +n^2 \sum_{i=1}^2\nabla_i e^v]+n^3 e^v.
\]
\end{proposition}

{\it Proof of Proposition \ref{phi-var}:}
We can now use our explicit evaluation of the Jacobi field $J$ (corresponding to 
the variation $\partial_\tau$ in the notation of \eqref{mod.Jac}) obtained in the previous 
subsection. In particular, our
first  aim is to 
determine the dependence of the quantities

\[
\partial_\tau|_{\tau=0} \mind^{AB}:=J(\mind^{AB}), 
\partial_\tau|_{\tau=0} \ul{\alpha}_{AB}:= J(\ul{\alpha}_{AB})
\]
on the function $v$. We will show that they depend as in  \eqref{eq.trulchi.var}, 
and we in fact
 determine the parameters in the RHSs of that equation. Then, Lemma
  \ref{trulchi.var} yields an explicit  formula for $\dot{a}_v(\la)$. 
\newline

We first calculate the first term above. We start by observing that: 
\[
J\mind^{AB}=-\sum_{C,D=1}^2J(\mind_{CD})\mind^{AC}\mind^{BD}.
\]
And then, recalling that $J\la=0$ and $e^A\la=0$ by construction, we calculate:

\[
\begin{split}
&J\mind_{AB}=Jg(e_A, e_B)=g(\nabla_Je_A,e_B)+g(e_A,\nabla_J e_B)=
\la^{-1}g(\nabla_J\partial_{\phi^A},e_B)
\\&+\la^{-1}
g(e^A,\nabla_J \partial_{\phi^B})
=\la^{-1}g(\nabla_{\phi^A}J,e_B)+\la^{-1}
g(e^A,\nabla_{\phi^B}J).
\end{split}
\]
To pursue this calculation, we express the Jacobi field $J$ in terms of the frame 
$\ul{L},L, e_1, e_2$: 
\[
J=J^{\ul{L}}\ul{L}+J^LL+\sum_{C=1}^2J^Ce_C.
\]

Thus: 

\beq
\label{Jmetric}
\begin{split}
&J\mind_{AB}=\la^{-1}\sum_{C=1}^2(\nabla_{\phi^A}J^C)g(e_C,e_B)+\la^{-1}
\sum_{C=1}^2(\nabla_{\phi^B}J^C)g(e_A,e_C)+\ul{\chi}_{AB}J^{\ul{L}}
+\chi_{AB}J^L
\\&+\sum_{C=1}^2J^C[g(\nabla_Ae_C, e_B)+g(e_A,\nabla_Be_C)].
\end{split}
\eeq
In view of \eqref{decay-conn}, and the formulas for the other components of $J$ that we found above, 
in conjunction with \eqref{decay-conn}, we derive that there exist functions 
$f^{AB}(\la)\in O^1_2(1)$, 
$r\in O^\delta_2(\la^{-1})$ so that:

\beq
\label{match}
\dot{\mind}^{AB}=J\mind_{AB}=f^{AB}(\la)[\mc{W}[e^v]+re^v],
\eeq
where $\mc{W}[e^v]$ stands for the RHS of \eqref{limit.J}.
Note that the above shows that the first assumption of Lemma \ref{trulchi.var} is fulfilled. 
\newline

Next, we calculate the variation of the curvature component $\ul{\alpha}_{AB}$.
We claim:

\begin{lemma}
\label{Conclusion}
 There exist functions $f_{AB}^1(\la), f_{AB}^2(\la)\in 
 O^\delta_2(\la^{-3-\epsilon})$, 
 $y_{AB}^C(\la) \in  O^\delta_2(\la^{-3-\epsilon})$
(independent of $v$) so that for each $A,B\in\{1,2\}$:
\beq
\begin{split}
&\dot{\ul{\alpha}}_{AB}=f^1_{AB}\{\Delta_{\mc{S}_0} e^v
-\rho e^v\}+f^2_{AB}e^v+y_{AB}^Ce_C(e^v).
\end{split}
\eeq
\end{lemma}

Observe that
the above implies that the second assumption of Lemma \ref{trulchi.var} is fulfilled. 
Thus, combining the above Lemma with \eqref{match}, and then invoking  Lemmas \ref{trulchi.var} and \ref{lumin.var}, 
  Proposition \ref{phi-var} follows immediately. Thus, matters are reduced to showing  Lemma \ref{Conclusion}.
\newline

{\it Proof:} We  calculate for each $A, B\in\{1,2\}$.

\beq
\begin{split}
\label{dot-alpha}
&\dot{\ul{\alpha}}_{AB}:=\partial_\tau|_{\tau=0} [\ul{\alpha}(e_A,e_B)]=
J[\ul{\alpha}(e_A,e_B)]=J[R(\ul{L}, e_A, e_B, \ul{L})]= (\nabla_J R)( \ul{L}, e_A, e_B, \ul{L})
\\&+R(\nabla_J\ul{L}, e_A, e_B, \ul{L})+R(\ul{L}, \nabla_J e_A, e_B, \ul{L})
+R(\ul{L}, e_A, \nabla_J e_B, \ul{L})+R(\ul{L}, e_A, e_B, \nabla_J\ul{L}).
\end{split}
\eeq
Observe that by our decay assumptions \eqref{curv.decay}, \eqref{L.curv.decay},
\eqref{ulL.curv.decay}
 on the derivatives of the Weyl curvature 
components, as well as the bounds \eqref{L.easy}, \eqref{solnform}, \eqref{limit.J}
 that we have derived, we find:
\beq
\begin{split}\label{name1}
&(\nabla_J R)(\ul{L}, e_A, e_B, \ul{L})=
J^{\ul{L}}\nabla_{\ul{L}}\ul{\alpha}_{AB}+J^{L}\nabla_{L}\ul{\alpha}_{AB}
+J^C\nabla_C\ul{\alpha}_{AB}
\\&=
O^\delta_2(\la^{-3-\epsilon})
\{\Delta_{\mc{S}} e^v- 2 \zeta^A\nabla_A 
e^v 
+[\frac{1}{4}tr\chi tr\ul{\chi}-\rho
+\frac{1}{2}\hat{\chi}\cdot\hat{\ul{\chi}}-div\zeta -|\zeta|^2][\mc{S}_0]e^v\}
\\&+O^\delta_2(\la^{-3-\epsilon})e^v
\end{split}
\eeq
In order to calculate the remaining terms in the RHS of \eqref{dot-alpha},
 it suffices to calculate the second and third terms in the RHS; the
 fourth and fifth follow in the same way.  We first consider the first term. 
By construction $J$ and $\ul{L}$ commute. 
Thus: 
\beq
\begin{split}
& R(\nabla_J\ul{L}, e_A, e_B, \ul{L})=R(\nabla_{\ul{L}}J, e_A, e_B, \ul{L})=
 (J^{\ul{L}})^\p R(\ul{L}, e_A, e_B, \ul{L})
\\&+\sum_{C=1}^2
 (J^C)^\p R(e_C, e_A, e_B,\ul{L})
 \\&+\sum_{C=1}^2J^C R(\nabla_{\ul{L}}e_C, e_A, e_B, \ul{L})+J^LR(\nabla_{\ul{L}}L,e_A, e_B, \ul{L})
\\&-\sum_{C=1}^2[\frac{J^C}{\la}R(e_C, e_A, e_B, \ul{L})+J^C\ul{\chi}_C^DR(e_D, e_A, e_B, \ul{L})+\zeta^CR(\ul{L}, e_A, e_B, \ul{L})
\end{split}
\eeq
(We have used the fact that $({J}^L)^\p=0$). Thus, using 
\eqref{curv.decay}, \eqref{L.curv.decay},
\eqref{ulL.curv.decay}, we derive that in the notation of Lemma \ref{Conclusion}:
\beq
\label{name2}
 R(\nabla_J\ul{L}, e_A, e_B, \ul{L})=(y^C_{AB}e_C)e^v+f^2_{AB}e^v,
\eeq
in the notation of Lemma \ref{Conclusion}. 
To evaluate the second term, note that since $J\la=0$, 
$[J, \partial_\phi]=[J,\partial_\theta]=0$  then $J$ and 
$e^A, A=1,2$ 
also commute. Thus, invoking \eqref{nablaL}, and  using the fact that 
$R(\ul{L},\ul{L}, e^B, \ul{L})=0$ we find:

\beq
\begin{split}
&R(\ul{L},\nabla_Je_A, e_B, \ul{L})=R(\ul{L},\nabla_{e_A}J, e_B, \ul{L})=
\sum_{C=1}^2(e_AJ^C)R(\ul{L}, 
 e_C, e_B,\ul{L})
\\&+(e_AJ^L)R(\ul{L},L,e_B,\ul{L})
\\&+J^{\ul{L}}R(\ul{L}, \nabla_{e_A}\ul{L}, e_B, \ul{L})
+J^{L}R(\ul{L}, \nabla_{e_A}L, e_B, \ul{L})+
\sum_{C=1}^2J^CR(\ul{L}, \nabla_{e_A}e_C, e_B, \ul{L}).
\end{split}
\eeq
Thus, using \eqref{dot-alpha}, the decay assumptions on the curvature coefficients 
 \eqref{curv.decay}, \eqref{L.curv.decay},
\eqref{ulL.curv.decay}, 
along with the 
bounds \eqref{L.easy}, \eqref{solnform}, \eqref{limit.J}
 we have obtained on the components of $J$ (in the previous subsection), 
we derive: 
\beq
\label{name3}
 R(\ul{L}, \nabla_J e_A, e_B, \ul{L})=(y^C_{AB}e_C)e^v+f^2_{AB}e^v;
\eeq
thus combining \eqref{name1}, \eqref{name2}, \eqref{name3} above,  Lemma 
\ref{Conclusion}
follows. This concludes the proof of Proposition \ref{phi-var}. $\Box$

\section{Finding the desired null hypersurface, via the implicit function theorem.}
\label{sec-inverse}

 We now recall Propositions  \ref{propo.var} and \ref{tJ-est}. Denote the RHS of
  \eqref{tJ-est-eqn} by $\mc{L}[e^v]$. We consider $\mc{L}$ acting on the space 
  $W^{4,p}(\mc{S}_0)$, for a fixed $p>2$.
\newline

We first carefully define the map to which we will apply the implicit function theorem.
Recall that  $\tau\cdot e^v=\omega$, and denote the 
sphere $\mc{S}_{v,\tau}\subset \mc{N}[\mc{S}_0]$ by $\mc{S}_\omega$. We have let
 $\lmcN_\omega:=\lmcN_{v,\tau}$, and 
 $\mc{S}^\p_\omega:=\mc{S}^\p_{v,\tau}\subset \lmcN_\omega$.
On each $\lmcN_\omega$, we recall the luminosity parameter $s_\omega$ with 
$\{s_\omega=1\}=\mc{S}'_\omega$. 

We define a natural map $\Psi:\lmcN_\omega\to\mc{S}_0$ which maps any point
 $Q\in \lmcN_\omega$ to the point $\Psi(Q)\in \mc{S}_0$ so that the 
 null generator of $\mc{N}[\mc{S}_0]$ though $\Psi(Q)$ and the null generator of 
 $\lmcN_\omega$ through $Q$ intersect (on $\mc{S}_\omega$)--see Figure 2 where now 
 $\mc{S}_{v,\tau}=\mc{S}_\omega$. 
 For conceptual convenience, let us consider the  ``boundary at infinity'' 
  $\partial_\infty \lmcN_\omega$ of $\lmcN_\omega$, which 
  inherits coordinates $\phi,
 \theta$ from $\mc{S}_0$ by the map $\Psi$.
We let   
 $\mc{S}^\infty_\omega:=\partial_\infty \lmcN_\omega$.

We then consider the operator:
$$\Phi^B:W^{4,p}(\mc{S}_0)\to L^p(\{ s=B\subset \lmcN_\omega\})$$  
 defined via:
 
 \beq
 \label{Phi-def}
 \Phi^B(\omega):=B^2\mc{K}(\{s_\omega=B\}\subset\lmcN_\omega)
 \eeq
 
 We also define:
  \beq
 \label{Phi-def-infty}
\Phi^\infty:W^{4,p}(\mc{S}_0)\to L^p(\mc{S}^\infty_\omega), 
 \Phi^\infty(\omega):=lim_{B\to\infty} \Phi^B(\omega)
 \eeq

We now claim that:

\begin{proposition}
\label{inverse-fn}
Choose any  $p>2$; the map $\Phi^\infty[\omega]:W^{4,p}(\mc{S}_0)\to 
L^p(\mc{S}^\infty_\omega)$, is well-defined and 
 $\mc{C}^1$ for all $\omega 
\in  \mc{B}(0,10^{-1}M^{-1})\subset W^{4,p}(\mc{S}_0)$, where $M(m)>1$ is a 
precise constant that will appear in the proof.\footnote{$\mc{B}(0, 10^{-1}M^{-1})$
stands for the ball of radius $10^{-1}M^{-1}$ in $W^{4,p}(\mc{S}_0)$.}
 Furthermore, letting 
\[
\dot{\Phi}^{\omega,\infty}:W^{4,p}(\mc{S}_\omega)\to L^p(\mc{S}^\infty_\omega)
\]
  be the linearization around any $\omega\in \mc{B}(0,10^{-1}M^{-1})\subset W^{4,p}
  (\mc{S}_0)$, 
   we claim that for 
  any   $\omega, \omega'\in \mc{B}(0,10^{-1}M^{-1})$, $\omega,\omega'\ge 0$ 
  we have:\footnote{Note that the operators 
  $\dot{\Phi}^{\omega,\infty}, \dot{\Phi}^{\omega^\p,\infty}$ take values over different 
  spheres, ($\mc{S}^\infty_\omega$ and $\mc{S}^\infty_{\omega^\p}$ 
  respectively). Nonetheless we 
  may compare and subtract the operators below via the natural map from these spheres 
  to $\mc{S}_0$ as described above. }
  \beq
  \label{operator-bound}
||  \dot{\Phi}^{\omega,\infty}- 
\dot{\Phi}^{\omega',\infty}||_{W^{4,p}(\mc{S}_\omega)\to L^p(\mc{S}^\infty)}\le 
M||\omega-\omega' ||_{W^{4,p}(\mc{S}_0)}.
  \eeq

Moreover, for $0<\delta<<m$ small enough\footnote{Recall that $\delta>0$ captures the closeness 
of the underlying space-time and $\lmcN$ to the ambient Schwarzschild space-time,
around a shear-free null surface. In particular recall that 
$||\mc{K}^\infty(\mc{N}_0)-1||_{W^{2,p}}$ is bounded by $\delta$.} 
we claim that there exists a constant $C>0$ with $|C-1|$ 
being small, and a function $\omega\in \mc{B}(0,10^{-1}M^{-1})\subset W^{4,p}(\mc{S}_0)$ 
so that:
\beq
\label{end-result}
\Phi^{\infty} (\omega)=C. 
\eeq
\end{proposition}

Note that the second part of  \eqref{end-result} implies Theorem \ref{the-result} above. 
As we will note in the proof, there are in fact many such $\omega$'s--in fact a 3-dimensional 
space of such functions. 
\newline

{\it Proof of Proposition \ref{inverse-fn}:} 
We will be showing this by an application of the implicit function theorem. Returning to
 expressing $\omega=\tau v$, 
we have derived, combining Lemma \ref{areas} with  
Propositions \ref{propo.var} \ref{tJ-est}, and the conventions on $\mc{L}[e^v]$ at the
 beginning of this section, that:
\beq
\label{finally}
\frac{d}{d\tau}|_{\tau=0}\Phi^\infty(e^v\tau):=\dot{\Phi}^\infty[e^v]=
(\Delta_{\gamma^\infty}+2\mc{K}[\gamma^\infty])\circ \mc{L}[e^v].
\eeq
In fact, as noted in Remark \ref{generalization} the same formula 
holds for the variation around any sphere $\mc{S}^\prime_\omega$, 
 for all $\omega(\mc{S}^2)\ge 0$
with $||\omega||_{\mc{C}^2}\le 10^{-1}m$. We denote the corresponding linearization 
as follows:
\beq
\label{finally'}
\frac{d}{d\tau}|_{\tau=0}\Phi^{\omega,\infty}(e^v\tau):=\dot{\Phi}^{\omega,\infty}[e^v]=
(\Delta_{\gamma^\infty_\omega}+2\mc{K}[\gamma^\infty_\omega])\circ 
\mc{L}_\omega[e^v].
\eeq
Observe that   the Sobolev embedding 
theorem $W^{4,p}(\mc{S}^2)\subset \mc{C}^3(\mc{S}^2)$ then implies that if we
 choose $M$ to be the norm of this embedding  times $m$, 
 then $\Phi^\infty[\omega]$ is a well-defined map for all 
$\omega\ge 0$, $\omega\in \mc{B}(0,10^{-1}M^{-1})$. 
Also \eqref{operator-bound}
follows by keeping track of the transformation laws for all the geometric quantities that 
appear as coefficients in the RHS of \eqref{finally}, along with the embedding 
$W^{4,p}(\mc{S}^2)\subset \mc{C}^3(\mc{S}^2)$. 
 Thus, the control of the modulus of continuity in 
\eqref{operator-bound} follows by keeping track of the transformation laws of these quantities 
under changes of $\omega$. 
We omit the details on these points, as they are fairly 
standard.  
\newline

To prove \eqref{end-result}, we need to understand 
the mapping properties of  $\dot{\Phi}^{\omega,\infty}$ for all  $\omega\in\mc{B}
(0,10^{-1}M^{-1})$: 

Observe that  for all $\omega\in\mc{C}^2(\mc{S}_0)$ 
$||\omega||_{\mc{C}^2(\mc{S}_0)}\le
10^{-1}m$ 
the operator $\mc{L}_\omega$ is a (non-self-adjoint) perturbation of the Laplacian
 $\Delta_{\mathbb{S}^2}$ on the round 2-sphere:
  \[
\mc{L}_\omega= \Delta_{\mathbb{S}^2} +\e^{ij}\partial_{ij}+\beta^i\partial_i+\eta. 
 \] 
 where the coefficients $\e^{ij}, \beta^i, \eta$ (all of which depend on $\omega$) 
 are small and bounded in the $\mc{C}^3$ norm. 
 This (by the continuous dependence of the 
 spectrum of second order elliptic operators on the operator coefficients) implies that 
 there exist  co-dimension-1 subspaces $\mc{A}_\omega\subset W^{4,p}(\mc{S}_0)$, 
 $\mc{T}_\omega\subset W^{2,p}(\mc{S}_0)$ so that the restriction  
 $\mc{L}^r_\omega$ of $\mc{L}$ 
 from $\mc{A}_\omega$ into 
 $\mc{T}_\omega$ is one-to-one and onto, and moreover 
 $\mc{L}_\omega$ is coercive,
\[
||\mc{L}_\omega^r(\psi)||_{W^{2,p}}\ge 10^{-1}m^{-1} ||\psi||_{\mc{A}_\omega}.
\] 
  Thus in particular the inverse $(\mc{L}^r_\omega)^{-1}:\mc{T}\to\mc{A}_\omega$ is  
 bounded:
 \[
 ||(\mc{L}_\omega^r)^{-1}\phi||_{W^{4,p}}\le 10m ||\phi||_{W^{2,p}}, \forall \phi\in
 \mc{T}_\omega.
 \]
Note  that 
 $\mc{A}_\omega$, $\mc{T}_\omega$ are annihilated by elements  $\psi_\omega$, 
 $\phi_\omega$  
 in the dual space which are $\delta$-close 
 (in the suitable norms)  to the constant function
  $1$ (this is because of the closeness of the operator $\mc{L}$ to 
  the round Laplacian $\Delta_{\mathbb{S}^2}$):
  \beq
  \label{space1}
\psi\in  \mc{A}_\omega\iff \{<\psi,\psi_\omega>=0 \text{ }\text{for some fixed}\text{ }
 \psi_\omega\in (W^{4,p})^*, 
||1-\psi_\omega||_{(W^{4,p})^*}\le \delta \}.
  \eeq
  
  \beq
  \label{space2}
  \psi\in  \mc{T}_\omega\iff \{<\psi,\phi_\omega>=0 \text{ }\text{for some fixed} \text{ }
  \phi_\omega\in (W^{2,p})^*, 
||1-\phi_\omega||_{(W^{2,p})^*}\le \delta \}.
  \eeq
  
  Now, observe that \eqref{finally'} implies that the effect on the Gauss curvature 
  of $\mc{S}^\infty_\omega$ under perturbations of $\lmcN_\omega$ (with a first variation 
  of $\omega$ by $e^v$)
  agrees with that induced by \emph{conformally} varying  the metric 
  $\gamma^\infty_\omega$, with $\mc{L}_\omega[e^v]$ being the first variation 
  of the conformal factor. In particular, by integrating over these variations, we derive that 
  $\Phi^\infty(\omega)$ (i.e.~the Gauss curvature   on
   $\mc{S}^\infty_\omega:=\partial_\infty\lmcN_\omega$) agrees with the Gauss curvature of  
   a metric 
  \[
  e^{2\Sigma(\omega)}\gamma^\infty,
  \]
where $\Sigma(\omega)$ (for $\omega=e^v$) is defined via:
\beq
\label{Sigma-define}
\Sigma(\omega)=\int_0^1 \mc{L}_{\tau\omega}[\omega]d\tau.
\eeq

In particular, the metric at infinity $\gamma^\infty_\omega$ on $\lmcN_\omega$ arises from the metric at infinity $\gamma^\infty$ on $\lmcN_0$ 
by multiplying against a conformal factor $e^{2\Sigma(\omega)}$, with $\Sigma(\omega)$ defined via \eqref{Sigma-define}.
Furthermore, note that 
 $\Sigma(\omega)$ is in $W^{2,p}(\mc{S}_0)$ (with $W^{2,p}$ norm bounded by 
  $10^{-1}mM$). Thus, it suffices to show that there exists an 
  $\omega\in W^{4,p}(\mc{S}_0)$ so that: 
  \beq
  \label{dhm1}
  \mc{K}[  e^{\Sigma(\omega)}\gamma^\infty]=C, 
  \eeq
for some constant $C\sim 1$. This will follow from the implicit function theorem. 
First, note that all the operators $\mc{L}_{\tau\omega}$ are small perturbations of 
$\Delta_{\mathbb{S}^2}$, as noted above. 
Next, to make the implicit function theorem applicable, we slightly generalize this by considering (for any $\mc{C}^3$-small function 
$\alpha$) the operator 
\[
\Sigma_\alpha[\omega]:=\int_0^1\mc{L}_{\tau \alpha}[\omega]d\tau,
\]
which is \emph{still} a small (non-self-adjoint) perturbation of the Laplacian; we are interested in the case where $\alpha=\omega$. 
Then, the implicit function theorem implies the following two facts:  First, together with the  bound on 
the modulus 
  of continuity of $\dot{\Phi}^{\omega,\infty}$ for all $\omega\in \mc{B}(0, 10^{-1}M^{-1})$ 
  implies that the image 
  $\Sigma(\mc{B}(0, 10^{-1}M^{-1}))$ is a codimension-1 smooth (open) submanifold
  in $W^{2,p}(\mc{S}_0)$, with radius bounded below by $100^{-1}M^{-1}$. 
  
  Secondly, using \eqref{space2}, it follows that 
  $\tilde{\Sigma}:=\Sigma(\mc{B}(1, 10^{-1}M^{-1}))$
  intersects the space of dilations (in $W^{2,p}$) transversely: Given any $P\in \tilde{\Sigma}$, considering the 
  segment ${\rm seg}(P):=t\cdot P$, $t\in [1-10\delta, 1+10\delta]$. Then the segment 
  ${\rm seg}(P)$ intersects $\tilde{\Sigma}$ only at $P$. Moreover, the intersection is
   ``almost normal'', in the sense that $\phi_P(P)\sim 1$, for $\phi$ being the element in the 
   dual space in \eqref{space2} that annihilates the tangent space of $\Sigma$ at $P$.
   
   Now, given the solution to the uniformization problem for the metric $\gamma^\infty$, note that there exists a 4-parameter space of functions $U_{\la, q}$
for which 
\[
\mc{K}[U_{\la,q}\gamma^\infty]=\la.
\]   
Here $\la\in \mathbb{R}_+$ and 
   $q\in SO(3,1)$ (the conformal group of $\mathbb{S}^2$), normalized so  that 
   $U_{\la,q}=\la^{-1} U_{1,q}$ (since the scaling of a conformal factor corresponds to a scaling 
   of the resulting Gauss curvature). Thus the space $U_{\la,q}$, 
   $\la\in\mathbb{R}_+, q\in SO(3,1)$ \emph{intersects} $\tilde{\Sigma}$ 
   \emph{transversely} along a smooth 3-dimensional submanifold. Choosing any element of this intersection provides a conformal factor $e^{\Sigma(\omega)}$ for \eqref{dhm1} to hold. 
   
   The three dimensions of freedom thus essentially correspond to the three dimensions of the conformal group modulo isometries, $SO(3,1)/SO(3)$. 
   They corresspond to the fact that we can capture all constant-curvature (up to the constant, which is not fixed here)
   metrics conformal to $\gamma^\infty$, nearby $\gamma^\infty$. This completes our proof. 
   $\Box$

  \end{document}